\documentclass[journal,10pt,twocolumn]{IEEEtran}

\usepackage{cite}
\usepackage{amssymb}
\usepackage{amsmath}
\usepackage{graphicx}
\usepackage{enumitem}
\usepackage{bm}
\usepackage{mathtools}
\usepackage{amsfonts}
\usepackage{stackengine}
\usepackage{xcolor}
\usepackage{stfloats}
\usepackage{array}
\newcolumntype{M}[1]{>{\centering\arraybackslash}m{#1}}
\newcolumntype{P}[1]{>{\centering\arraybackslash}p{#1}}
\stackMath
\usepackage{multirow} 
\usepackage{framed}
\usepackage{fancyhdr,lipsum}
\fancypagestyle{mahmood}{%
	\fancyhf{} 
	
	\fancyhead[L]{\scriptsize THIS WORK HAS BEEN SUBMITTED TO THE IEEE FOR POSSIBLE PUBLICATION. COPYRIGHT MAY BE TRANSFERRED WITHOUT NOTICE.}
}%
\makeatletter
\let\ps@IEEEtitlepagestyle\ps@mahmood
\makeatother

\newcommand{\kernel}{{\ooalign{$k$\cr\raisebox{0.2em}{\kern0.08em--}\cr}
}}
\DeclareRobustCommand{\mhl}[1]{%
	\ifmmode\text{\color{black}{$#1$}}\else{\color{black}{#1}}\fi
}
\makeatletter
\newcommand{\vast}{\bBigg@{3}}
\makeatother

\makeatletter
\@ifdefinable\@latex@chi{\let\@latex@chi\chi}
\renewcommand*\chi{{\@latex@chi\smash[t]{\mathstrut}}} 
\makeatletter
\def\mathclap#1{\text{\hbox to 0pt{\hss$\mathsurround=0pt#1$\hss}}}

\makeatother
\def\mathclap#1{\text{\hbox to 0pt{\hss$\mathsurround=0pt#1$\hss}}}

\usepackage[font=footnotesize]{caption}
\usepackage[font=footnotesize]{subcaption} 

\usepackage{relsize}
\usepackage{bm}
\usepackage{soul}
\usepackage{empheq}
\usepackage{placeins}

\begin{document}

	\title{A New Correction to the Rytov Approximation for Strongly Scattering Lossy Media}
	
\author{Amartansh~Dubey,~\IEEEmembership{Graduate Student Member,~IEEE}, Xudong~Chen,~\IEEEmembership{Fellow,~IEEE} and~Ross~Murch,~\IEEEmembership{Fellow,~IEEE}\thanks{This work was supported by the Hong Kong Research Grants Council with the General Research Fund grant 16211618 and the Collaborative Research Fund C6012-20G.}\thanks{A. Dubey is with the Department of Electronic and Computer Engineering, Hong Kong University of Science and Technology (HKUST), Hong Kong, (e-mail:
		\protect{}{adubey@connect.ust.hk})}.
	\thanks{X. Chen is with the Department of Electrical and Computer Engineering, National University of Singapore, Singapore.} \thanks{R. Murch is with the Department of Electronic and Computer Engineering
		and the Institute of Advanced Study both at the Hong Kong University
		of Science and Technology (HKUST), Hong Kong.}}

	\maketitle

\begin{abstract}
We propose a correction to the conventional Rytov approximation (RA) and investigate its performance for predicting wave scattering under strong scattering conditions. An important motivation for the correction and investigation is to help in the development of better models for inverse scattering. The correction is based upon incorporating the high frequency theory of inhomogeneous wave propagation for lossy media into the RA formulation. We denote the technique as the extended Rytov approximation for lossy media (xRA-LM). xRA-LM significantly improves upon existing non-iterative linear scattering approximations such as RA and the Born approximation (BA) by providing a validity range for the permittivity of the objects of up to 50 times greater than RA. We demonstrate the technique by providing results for predicting wave scattering from piece-wise homogeneous scatterers in a two-dimensional (2D) region. Numerical investigation of the performance of xRA-LM for solving direct problem show that xRA-LM can accurately predict wave scattering by electrically large, low-loss scatterers with high complex permittivity ($\epsilon_r> 50+5j$). To the best of our knowledge, this is the first non-iterative, linear approximate wave scattering model which has a large validity range in terms of both permittivity and electrical size.
\end{abstract}

\IEEEpeerreviewmaketitle
	
\section{Introduction}
\label{Sec_Intro}
Modeling of electromagnetic wave scattering has provided many technological breakthroughs related to the solution of direct and inverse problems \cite{chen2018computational, jones1986acoustic, MoMoments, FTDTref, Mittra1998, chen2010subspace, murch1990inverse, benny2020overview, bates1991manageable, nikolova2017, DubeyTxline, jing2018approximate, balanis2015antenna, Dubey2021, deshmukh2021physics, oristaglio1985accuracy}. Important applications have included indoor propagation prediction, microwave imaging (inverse scattering) and antenna design. The exact solution of electromagnetic wave scattering can be found from Maxwell's equations using formulations such as the volume source integral (VSI) (Lipmann-Schwinger equation) or Eigenfunction expansions \cite{chen2018computational, murch1990inverse, jones1986acoustic}. These exact formulations can be solved numerically using techniques such as the Method of Moments (MoM), and the Finite-Difference Time-Domain (FDTD) \cite{MoMoments, FTDTref, Mittra1998, chen2018computational, taflove2005computational} and these have revolutionized the design of antennas and radio frequency (RF) circuits in recent decades by providing very accurate predictions of wave scattering and radiation. 

However, for applications where the domain of interest (DoI) is electrically large or contains very high permittivity materials, solving exact models as a direct problem becomes computationally infeasible, and as inverse problem the exact models become highly non-linear and ill-posed and cannot handle the measurement data suffering from inaccuracies due to noise and real world data acquisition process \cite{chen2018computational, chen2010subspace, jones1986acoustic, DubeyTxline, jing2018approximate, nikolova2017, bates1991manageable}. In practice, such conditions involving an electrically large DoI with strong scatterers and imperfect measurements are encountered in many application such as indoor imaging,  non-destructive evaluation, and microwave imaging  \cite{chen2018computational, chen2010subspace, jones1986acoustic, DubeyTxline, jing2018approximate, nikolova2017, bates1991manageable, depatla2015x, chen2020review, Patwari2010}. Under these conditions, the exact models (such as VSI) have found limited practical application. This opens up a huge research field for finding simpler approximations to these exact models which can be solved with feasible computational requirements and practical measurement systems.

Common approximate models include the Born Approximation (BA), Rytov approximation (RA), Geometrical Optics (GO), and Uniform Theory of Diffraction (UTD) \cite{chen2018computational, bornwolf, 6324957, jones1986acoustic, chen2010subspace, murch1990inverse, jones1970ray, benny2020overview, bates1991manageable, nikolova2017, wu2003wave, enright1992towards, habashy1993beyond, murch1992extended, oristaglio1985accuracy}. Many of these approximate techniques have provided the basis for well known approximate inverse scattering techniques and for extending wave scattering theory to practical applications \cite{chen2018computational, jones1986acoustic, murch1990inverse, benny2020overview, DubeyTxline, chen2010subspace, jing2018approximate, nikolova2017, bates1991manageable, depatla2015x, 9107142, bose2007, 7029113, 6165627, 4120260, 704854}. Among the approximate techniques, the computationally least complex include the \textit{non-iterative linear approximations} such as RA and BA and these are commonly used in inverse scattering \cite{chen2018computational, bornwolf, jones1986acoustic, wu2003wave, bates1991manageable, oristaglio1985accuracy}. While these are useful, they have a limited range of validity. For example, BA fails if the scatterer is electrically large or has permittivity deviating significantly from unity while RA fails when permittivity deviates significantly from unity.

In this work, we propose a non-iterative linear approximation which can estimate scattering from strongly scattering objects  with very high relative permittivity (up to $\epsilon_r> 50+5j$) and large electrical size (greater than the incident wavelength). Our technique is based upon a correction to conventional RA and we denote the technique as the extended Rytov approximation for lossy media (xRA-LM). An important motivation for the correction and investigation is to help in the development of better models for inverse scattering. xRA-LM incorporates the high frequency theory of inhomogeneous wave propagation for lossy media \cite{yang2009effective, zhang2020generalized, zhang2015refractive, chang2005ray, yang1995light, jones1970ray} into the formulation of RA, resulting in a remarkably higher validity range. Another key aspect of xRA-LM is its validity for lossy media and this is important for applications in the everyday environment where most materials have a loss component.  For example, the complex-valued relative permittivity, $\epsilon_r = \epsilon_R  + j\epsilon_I$, of scatterers in the everyday environment at 2.4 GHz have $\epsilon_R$ ranging from $2< \epsilon_R \le 50$ (where $\epsilon_R \ge 20$ for the human body for example) \cite{4562803, Productnote, ahmad2014partially, Dubey2021}) and $\epsilon_I$ as characterized by loss tangent is in the range $\delta = \epsilon_I/\epsilon_R \in [10^{-3}, 10^{-1}]$ \cite{4562803, Productnote, ahmad2014partially, Dubey2021}. 

We numerically investigate the performance of xRA-LM for solving direct problems by providing results of wave scattering from a 2D region. The results show that xRA-LM can accurately predict wave scattering from electrically large scatterers with large complex permittivity ($\epsilon_r> 50+5j$). Comparisons with RA and BA also show that xRA-LM significantly outperforms RA and BA while maintaining similar computational complexity. To the best of our knowledge, this is the first non-iterative, linear approximate wave scattering model which has such a high validity range (in terms of permittivity and size of scatterer). It can open up a new paradigm of non-iterative linear models that provide practically feasible solutions to both direct and inverse scattering problems in strong scattering environments \cite{chen2018computational, jones1986acoustic, murch1990inverse, benny2020overview, DubeyTxline, chen2010subspace, jing2018approximate, nikolova2017, bates1991manageable, depatla2015x, 9107142, bose2007, 7029113, 6165627, 4120260, 704854}. For example, a simplified phaseless version of xRA-LM has been utilized for inverse scattering to obtain impressive reconstruction results \cite{dubey2021accurate}. However the analysis of the accuracy of the underlying direct problem can provide a more comprehensive approach to analyzing the model. Unlike the inverse problem, the direct problem is not ill-posed and hence the accuracy obtained is directly related to the accuracy of the model rather than the regularization methods deployed to tackle ill-posedness. The goal of this paper is to investigate xRA-LM from the direct problem perspective to analyze its performance and accuracy.

\textit{Organization of paper:} The problem formulation is described in Section \ref{Sec_PF} followed by derivation of the proposed xRA-LM approximation in Section \ref{Sec_RytovOptics}. Section \ref{Sec_results} provides numerical results followed by Conclusions. In the remainder of this paper, lower and upper case boldfaced letters are used to represent vectors and matrices respectively. Italic letters are used to represent scalar quantities.

\section{Problem Formulation}
\label{Sec_PF}
\begin{figure}[!h]
	\centering
	\includegraphics[width=2in]{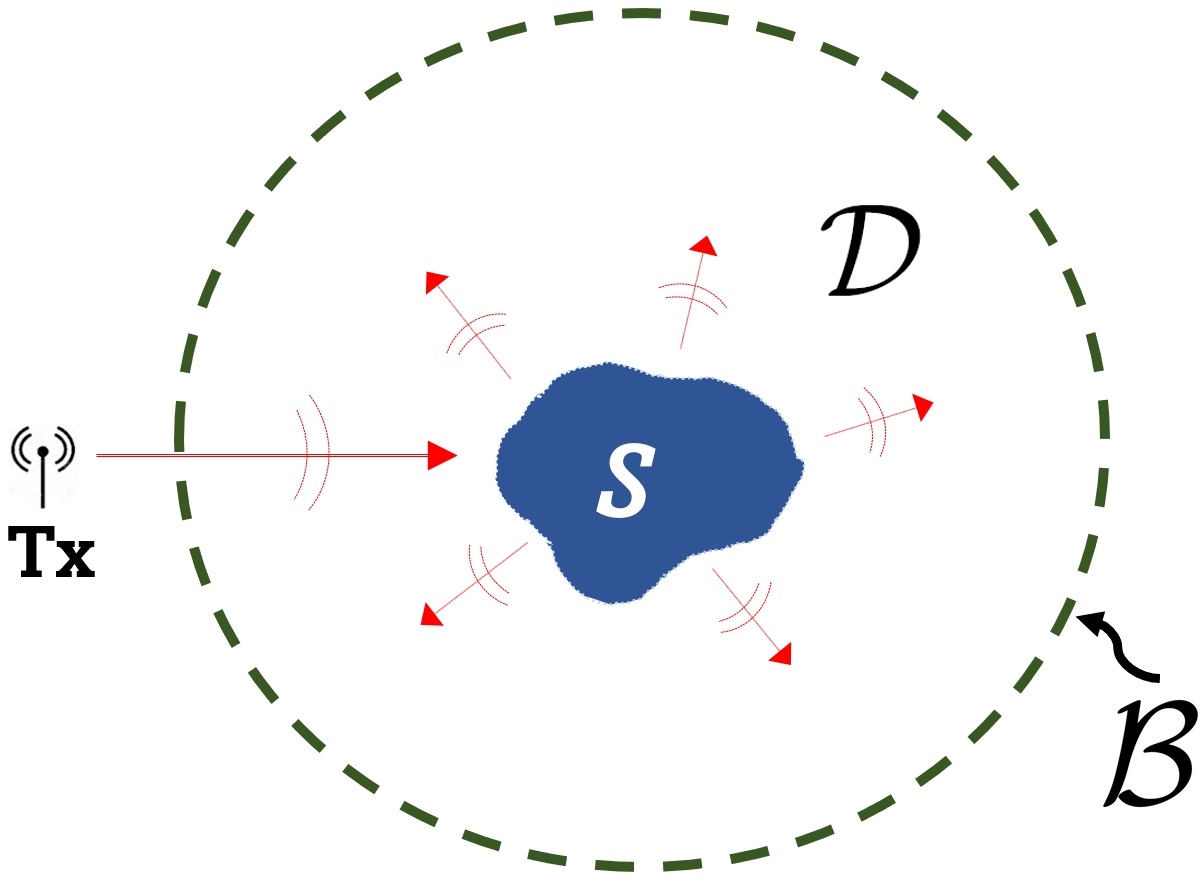}
	\caption{Illustration of wave scattering by a scattering object $\mathcal{{S}}$ placed inside the DOI $\mathcal{D}$. The receivers are placed along a circular boundary $\mathcal{B}$.}
	\label{geometry}
	\vspace{-0.2\baselineskip}
\end{figure}

Consider the 2D scenario shown in Fig. \ref{geometry} where a transmitter \textbf{Tx} or source of electromagnetic radiation illuminates an arbitrary shaped scatterer $\mathcal{S}$ placed inside DoI, $\mathcal{D}$. The scatterer is characterized by its complex-valued permittivity $\epsilon_r(\bm{r}) = \epsilon_R(\bm{r})+j \epsilon_I(\bm{r})$ (assuming permeability to be $\mu_0=1$) and the scattering from the DOI is collected by an array of receivers placed around a measurement boundary $\mathcal{B}$. The electromagnetic radiation from Tx is assumed to be monochromatic, time harmonic and vertically polarized which is often referred to as transverse magnetic (TM) in wave scattering context and is commonly used in real-world applications.

In the absence of any scatterers, the incident field at any point inside $\mathcal{D}$ is denoted by $E_i(\bm{r})$. It satisfies the free-space wave equation,
\begin{equation}
	\label{Eq_Hzfree}
	\begin{aligned}
		(\nabla^2 + k_0^2) E_i(\bm{r}) = 0, \qquad && \bm{r} \in \mathcal{D}
	\end{aligned}
\end{equation}
where $k_0=2\pi/\lambda_0$ is the free-space wavenumber and $\lambda_0$ is the free-space wavelength. The total field at any point is the sum of the incident and scattered field (i.e. $E(\bm{r}) = E_i(\bm{r}) + E_s(\bm{r}), \ \bm{r} \in \mathcal{D}$) and satisfies the inhomogeneous Helmholtz wave equation, 
\begin{equation}
	\label{Eq_Hz}
	\begin{aligned}
		(\nabla^2 + k_0^2 \nu^2(\bm{r})) E(\bm{r}) = 0, \quad && \bm{r} \in \mathcal{D}
	\end{aligned}
\end{equation}
where $\nu(\bm{r}))$ is refractive index and is related to relative permittivity through $\epsilon_r(\bm{r})) = \nu^2(\bm{r}))$. Subtracting (\ref{Eq_Hzfree}) and (\ref{Eq_Hz}) provides a wave equation in terms of scattered field $E_s = E-E_i$, and written as a Fredholm integral equation of the second kind,
\begin{equation}
	\label{Eq_VSI}
	\begin{aligned}
		E(\bm{r}) = E_i(\bm{r}) + k_0^2 \int_{\mathcal{D}} g(\bm{r}, \bm{r}') (\nu^2(\bm{r})-1) E(\bm{r}') d\bm{r}' 
	\end{aligned}
\end{equation}
where, $r \in \mathcal{B}, r' \in \mathcal{D}$. Equation (\ref{Eq_VSI}) is also known as the Lipmann-Schwinger equation or VSI and provides an exact description for wave scattering \cite{Mittra1998, chen2018computational}.

Solving VSI as a \textit{direct problem} implies estimation of $E(\bm{r})$ given $\nu(\bm{r})$ and $E_i(\bm{r})$. This is computationally expensive when $\mathcal{D}$ is electrically large because it requires estimation of $E(\bm{r})$ over $\mathcal{D}$. Solving VSI as an \textit{inverse problem} implies solving it for $\nu(\bm{r})$ given $E(\bm{r})$ and $E_i(\bm{r})$ on $\mathcal{B}$ which results in a non-linear, ill-posed problem since both $\nu(\bm{r})$ and  $E(\bm{r})$ need to be found inside $\mathcal{D}$. To overcome these challenges non-iterative linear approximations have been proposed to simplify VSI. The most extensively researched techniques are RA and BA. BA straightforwardly approximates the total field $E(\bm{r})$ inside the integral (\ref{Eq_VSI}) by the incident field $E_i(\bm{r})$ to give 
\begin{equation}
	\label{Eq_BA}
	\begin{aligned}
		E(\bm{r}) = E_i(\bm{r}) + k_0^2 \int_{\mathcal{D}} g(\bm{r}, \bm{r}') (\nu^2(\bm{r})-1) E_i(\bm{r}') d\bm{r}' ,
	\end{aligned}
\end{equation}
where, $\bm{r} \in \mathcal{B}$ and $\bm{r}' \in \mathcal{D}$. Therefore for the direct problem, unlike VSI, BA does not have any unknown field inside the integral, removing the need for any expensive matrix inversion step. It is also linear as an inverse problem (unlike VSI). However, BA has a poor range of validity as it fails for even a small permittivity contrast or if the size of the scatterer is comparable or larger than $\lambda_0$ \cite{chen2018computational, bornwolf, jones1986acoustic, wu2003wave, bates1991manageable, oristaglio1985accuracy}. 

RA on the other hand utilizes the Rytov transform to arrive at an approximate method that can handle electrically large objects but with a similar range of validity on permittivity as BA. The Rytov transformation normalizes the total field $E(\bm{r})$ by the incident field $E_i(\bm{r})$ to express the scattering by a complex phase $\phi_s(\bm{r})$,
\begin{equation}
	\label{Eq_rytTfield}
	\begin{aligned}
		\frac{E(\bm{r})}{E_i(\bm{r})} &= e^{\phi_s(\bm{r}) }.
	\end{aligned}
\end{equation}
Intuitively, the complex phase $\phi_s(\bm{r})$ represents the phase and log amplitude deviations from the incident field (caused by scattering). Substituting (\ref{Eq_rytTfield}) in (\ref{Eq_Hz}) and using (\ref{Eq_Hzfree}) gives a non-linear differential equation (Riccati equation in $E_i \phi_s$) \cite{wu2003wave},
\begin{equation}
		\label{Eq_rytdiffeq3}
		\begin{aligned}
			(\nabla^2 +  k_0^2) & (E_i(\bm{r}) \phi_s(\bm{r})) =\\ &-k_0^2 E_i(\bm{r}) \big[\nu^2(\bm{r})-1 + \frac{\nabla \phi_s(\bm{r}) \cdot \nabla \phi_s(\bm{r})}{k_0^2}\big].
		\end{aligned}
\end{equation}
Equation (\ref{Eq_rytdiffeq3}) can be written in integral form (which we call the Rytov integral (RI)) to obtain an expression for total field,
\begin{subequations}
 	\label{Eq_rytov2}
 	\begin{align}
 		E(\bm{r})  &= E_i(\bm{r})  \exp\bigg (  \frac{k_0^2}{E_i(\bm{r})} \int_{\mathcal{D}}  g(\bm{r}, \bm{r'})  \chi_{\text{RI}}(\bm{r}') E_i(\bm{r'}) d\bm{r'}^2\bigg),\\
	\chi_{\text{RI}} & (\bm{r}') = \nu(\bm{r'})^2-1 + \frac{\nabla \phi_s(\bm{r}) \cdot \nabla \phi_s(\bm{r})}{k_0^2},
	\end{align} 
\end{subequations}
where $\chi_{\text{RI}}$ is the contrast function of RI. The term $\nabla \phi_s \cdot \nabla \phi_s$ in RI is then neglected under a weak scattering assumption to arrive at RA as,
\begin{subequations}
	\label{Eq_RA}
	\begin{align}
		&E(\bm{r})  = E_i(\bm{r})  \exp\biggr(  \frac{k_0^2}{E_i(\bm{r})} \int_{\mathcal{D}}  g(\bm{r}, \bm{r'})  \chi_{\text{RA}}(\bm{r}') E_i(\bm{r'}) d\bm{r'}^2\biggl),\\
		& \qquad \qquad \qquad \chi_{\text{RA}}(\bm{r}')  = \nu(\bm{r'})^2-1.
	\end{align} 
\end{subequations}
Neglecting $\nabla \phi_s\cdot \nabla \phi_s$, makes RA useful only for weak scattering with $\epsilon_R \approx 1$ (similar to BA). However, RA does not impose a restriction on the size of the scatterer unlike BA \cite{wu2003wave}. For high permittivity variations, $\nabla \phi_s \cdot \nabla \phi_s$ cannot be neglected and estimation of $\nabla \phi_s \cdot \nabla \phi_s$ is difficult as it requires solving the intractable non-linear equation (\ref{Eq_rytov2}) \cite{caorsi1996rytov, nikolova2017}. To the best of our knowledge this has not been done for strongly scattering lossy media. Therefore, approximating $\nabla \phi_s\cdot \nabla \phi_s$, instead of completely neglecting it, can provide improvement over conventional RA.

\section{Corrections to the Rytov Approximation}
\label{Sec_RytovOptics}
In this section, we derive the proposed xRA-LM technique by providing corrections to RA using the characterization of inhomogeneous waves in lossy media \cite{yang2009effective, zhang2020generalized, zhang2015refractive, chang2005ray, yang1995light, jones1970ray}.

\subsection{High Frequency Approximations in Lossy Media}
\label{Sec_rayoptics}
High frequency approximations treat waves as straight rays and are used in approximations such as GO and GTD to describe scattering from objects that are larger in size compared to $\lambda_0$. However, ray formulations inside lossy media (with complex-valued refractive index) are intricate as the waves become inhomogeneous inside lossy media \cite{yang2009effective, zhang2020generalized, zhang2015refractive, chang2005ray, yang1995light, jones1970ray}. Due to this, there are surprisingly limited ray formulations for lossy media \cite{bornwolf, jones1970ray} even after decades of research. Inhomogeneous waves exhibit the property that the \textit{planes of constant phase} are no longer parallel to the \textit{planes of constant amplitude} \cite{yang2009effective, zhang2020generalized, zhang2015refractive, chang2005ray, yang1995light, jones1970ray}. In this work, we deal with homogeneous plane waves (HPW) that are incident on lossy media and become inhomogeneous plane waves (IPW) inside the lossy media.
 
\begin{figure}[!h]
	\centering
	\includegraphics[width=2.4in]{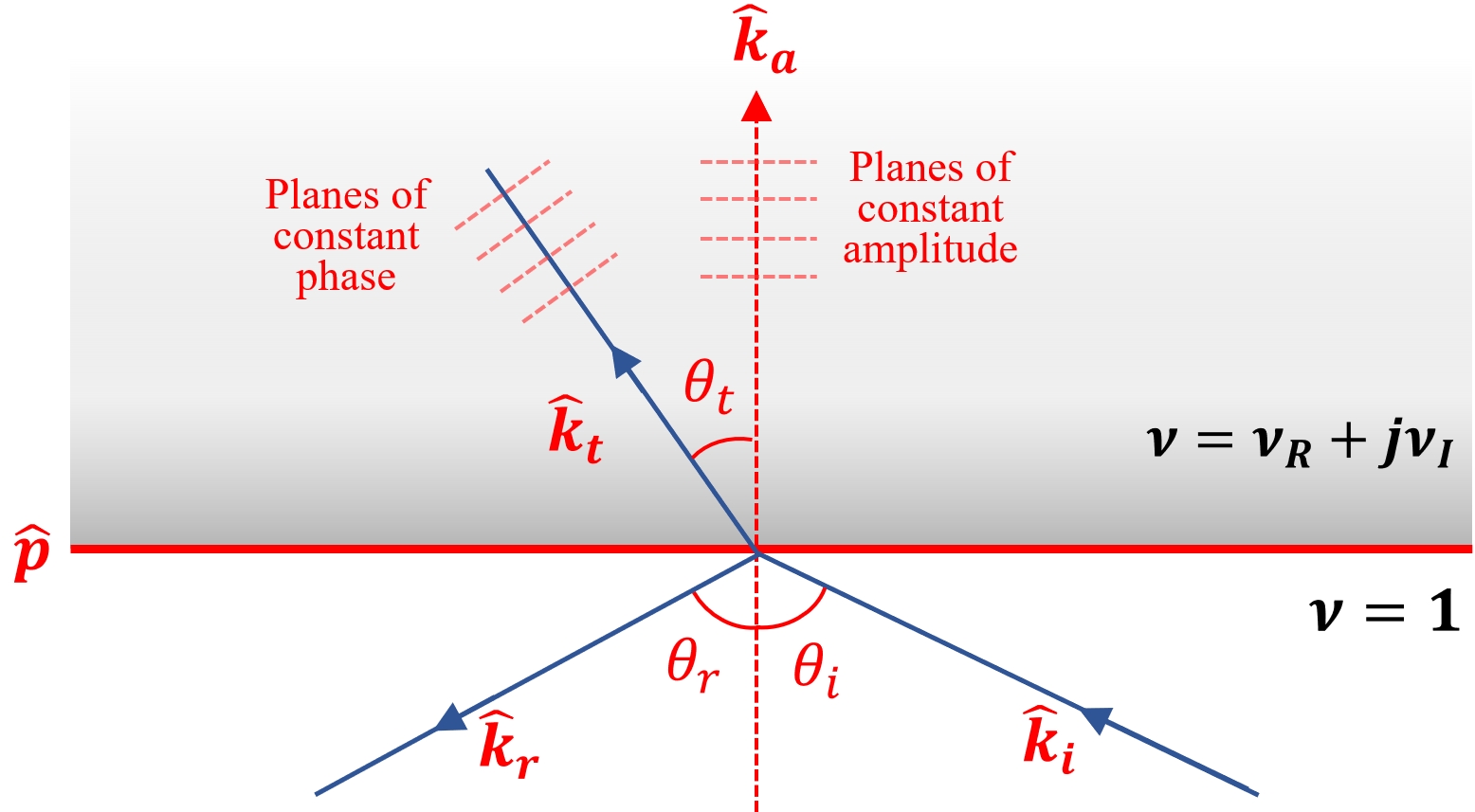}
	\caption{Free space to lossy media interface. The homogeneous plane wave (HPW) in lossless media becomes inhomogeneous plane wave (IPW) inside the lossy media (with refractive index $\nu = \nu_R+ j \nu_I$).}
	\label{interface}
	\vspace{-0.3\baselineskip}
\end{figure}
Fig. \ref{interface} illustrates a vacuum/air to lossy dielectric interface. The lossy media is characterized by a constant complex refractive  
index $\nu$ and constant relative permittivity $(\epsilon_r)$ \cite{bornwolf} defined by,
\begin{equation}
	\label{Eq_complexNu}
	\begin{aligned}
		\nu = \nu_R + j \nu_I; \quad
		\epsilon_r = \epsilon_R + j \epsilon_I,
	\end{aligned}
\end{equation}
and are related as $\nu^2 = \epsilon_r$ \cite{bornwolf}. Equating real and imaginary parts gives relations $	\nu_R^2 - \nu_I^2 = \epsilon_R $ and $
\nu_R \nu_I = \epsilon_I/2$. Using this, and defining loss tangent of the medium as $\delta = \epsilon_I/\epsilon_R$, we can express $\nu_R$ and $\nu_I$ as,
\begin{equation}
	\label{Eq_complexNu_R}
	\nu_R = \sqrt{\frac{\epsilon_R(\sqrt{1+\delta^2} + 1)}{2}}, \ \ \nu_I = \sqrt{\frac{\epsilon_R(\sqrt{1+\delta^2} - 1)}{2}}.
\end{equation}
For low-loss media ($\delta \ll 1$), (\ref{Eq_complexNu_R}) can be simplified (for practical use) using the binomial expansion as,
\begin{subequations}
	\label{Eq_complexNu_R_LL}
	\begin{align}
		\nu_R & \approx\sqrt{\epsilon_R + {\frac{1}{4}  \delta^2 \epsilon_R} } \approx \sqrt{\epsilon_R}\ ,\\
		\nu_I  & \approx \frac{\epsilon_I}{2 \sqrt{\epsilon_R} } = \frac{1}{2} \delta \sqrt{\epsilon_R} \ .
	\end{align}
\end{subequations} 
When the HPW field is incident on the air-lossy media interface (in Fig. \ref{interface}), it gets partially reflected (as HPW) and transmitted (as IPW) at the interface. Snell's law $(\sin\theta_i = (\nu_R + j\nu_I)\sin\theta_t)$ shows that the angle of refraction becomes a complex quantity in this case. Such complex angles are not geometrically intuitive in conventional GO and hence, a new concept of effective refractive index has been introduced \cite{yang1995light, yang2009effective, zhang2020generalized} which can be used to remove complex angles and is also used in our work. The concept of effective refractive index allows us to decompose the mathematical form of IPW as a linear combination of vectors normal to the planes of constant phase and constant amplitude. Then, real-valued Snell's law angle can be applied separately for the refraction and attenuation components of IPW (see \cite{chang2005ray, zhang2015refractive, zhang2020generalized, yang1995light, yang2009effective}). Using this concept, we can express the wave-vector of transmitted IPW field inside the lossy medium as,
\begin{equation}
	\label{Eq_complexwavevector}
	\begin{aligned}
		\bm{k}_t  = k_0 (V_R \bm{\hat{k}_t} + j V_I \bm{\hat{k}_a})
	\end{aligned}
\end{equation}
where, as shown in Fig. \ref{interface}, the unit vectors $\bm{\hat{k}_t}$ and $\bm{\hat{k}_a}$ are normal vectors respectively to the planes of constant phase and constant amplitude. $V_R$ and $V_I$ are the scalar coefficients of these unit vectors and are called effective real and imaginary parts of the refractive index respectively \cite{chang2005ray, zhang2015refractive, zhang2020generalized, yang1995light, yang2009effective}.  

In the free-space half of the interface, there will be two HPW fields, namely incident and reflected fields which can be written as,
\begin{subequations}
	\label{Eq_incidentwavevectorfield}
			\begin{align}
		E_i(\bm{r}) &= A_0 \exp{(jk_0 \ \bm{\hat{k}}_i \cdot \bm{r})}\\
		E_r(\bm{r}) &= A_r\exp{(jk_0 \ \bm{\hat{k}}_r \cdot \bm{r})} 
				\end{align}
\end{subequations}
where $\bm{\hat{k}}_i$ and $\bm{\hat{k}}_r$ are the wave-vectors of incident and reflected fields respectively. Further, using (\ref{Eq_complexwavevector}), we can express the transmitted IPW field inside the lossy medium as,
	\begin{equation}
		\label{Eq_complexwavevectorfield}
		\begin{aligned}
			E_t(\bm{r}) &= A_t\exp{(jk_0 \ (V_R \bm{\hat{k}_t} \cdot \bm{r}  + jV_I \bm{\hat{k}_a} \cdot \bm{r}))}
		\end{aligned}
	\end{equation}
We know that the phase of the incident, reflected and transmitted fields should match tangentially at the media interface, which gives,
\begin{equation}
	\label{Eq_SnellLaw}
	\begin{aligned}
		\bm{\hat{p}} \cdot \bm{\hat{k}_i} = 	\bm{\hat{p}} \cdot \bm{\hat{k}_r} = \bm{\hat{p}} \cdot	(V_R \bm{\hat{k}_t} + j V_I \bm{\hat{k}_a} ) 
	\end{aligned}
\end{equation}
By equating real parts of (\ref{Eq_SnellLaw}) gives real-valued Snell's law angles for the interface,
\begin{equation}
	\label{Eq_SnellLaw1}
	\begin{aligned}
		\bm{\hat{p}} \cdot \bm{\hat{k}_i} = \bm{\hat{p}} \cdot \bm{\hat{k}_r} = \bm{\hat{p}} \cdot	(V_R \bm{\hat{k}_t} )\\
		\sin\theta_i = \sin\theta_r = V_R \sin\theta_t.
	\end{aligned}
\end{equation}
and on equating imaginary parts of (\ref{Eq_SnellLaw}) shows that the vector $\bm \hat{k}_a$ is normal to the interface, i.e.,
\begin{equation}
	\label{Eq_SnellLaw2}
	\begin{aligned}
		\bm{\hat{p}} \cdot	(V_I\bm{\hat{k}_a} ) = 0  \implies \bm{\hat{p}} \perp \bm{\hat{k}_a}.
	\end{aligned}
\end{equation}
This is an interesting result as it implies that whenever a HPW field enters a lossy media and becomes IPW field, the planes of constant amplitude becomes parallel to the interface. This result is key to our derivation of xRA-LM as we shall show later. 

In the above results, it can be seen that there are no complex angles due to the use of effective refractive index expressions (\ref{Eq_complexwavevectorfield}). The effective refractive index ($V_R, V_I$) has to be related to the actual refractive index ($\nu_R, \nu_I$) so that it can be used for scattering estimation in xRA-LM. This can be performed by substituting (\ref{Eq_complexwavevectorfield}) into (\ref{Eq_Hz}) which gives 
	\begin{subequations}
		\label{Eq_effective}
		\begin{align}
			V_R^2 - V_I^2 = \nu_R^2 - \nu_I^2 \\
			V_R {V_I}\cos\theta_t 	= \nu_R \nu_I
		\end{align}
	\end{subequations}
where, from Fig. \ref{interface}, $\bm{\hat{k}_t} \cdot \bm{\hat{k}_a}= \cos\theta_t$. $V_R$ can be estimated by eliminating $V_I$ from (\ref{Eq_effective}) which gives, 
\begin{equation}
	\label{Eq_effectivefinal}
	\begin{aligned}
		V_R &= \bigg\{\frac{1}{2} \bigg(\sqrt{(\nu_R^2 - \nu_I^2)^2 +  4 \bigg[\frac{ \nu_R \nu_I}{\cos\theta_t}\bigg]^2} + \nu_R^2-\nu_I^2  \bigg) \bigg\}^{1/2}, 
	\end{aligned}
\end{equation}
where, $\cos\theta_t$ can be expressed in terms of $\sin\theta_i$ using (\ref{Eq_SnellLaw1}). Similarly, $V_I$ can be obtained by eliminating $V_R $ from (\ref{Eq_effective}).

Under the low-loss assumption $(\epsilon_R\gg \epsilon_I)$, (\ref{Eq_effectivefinal}) can be simplified (using binomial expansion \cite{chang2005ray}) to approximate $V_R$ as, 
\begin{equation}
	\label{Eq_VRVIlowloss1} 
	\begin{aligned}
		V_R & \approx \nu_R \bigg(1+\frac{\sin^2\theta_i}{2(\nu_R^2 - \sin^2\theta_i)} \delta^2 \bigg) \approx \nu_R
	\end{aligned}
\end{equation}
Similarly, $V_I$ can be approximated as,
\begin{equation}
	\label{Eq_VRVIlowloss2}
	\begin{aligned}
		V_I &\approx \frac{\nu_R \nu_I}{\sqrt{\nu_R^2 - \sin^2\theta_i}}\bigg(1-\frac{\nu_R^2 \sin^2\theta_i}{2(\nu_R^2-\sin^2\theta_i)}\delta^2 \bigg) \\
		& \approx \frac{\nu_R \nu_I}{\sqrt{\nu_R^2 - \sin^2\theta_i}} 
	\end{aligned}
\end{equation}


Using these results, we can rewrite the ray formulation (\ref{Eq_complexwavevectorfield}) inside an extended lossy scatterer with a piece-wise homogeneous distribution of complex-valued refractive index. This can be performed by rewriting (\ref{Eq_complexwavevectorfield}) using the path integral along the ray direction  ($\bm{{dr}} =dr \ \bm{\hat{k}_t}$) as,
\begin{equation}
	\label{Eq_raytraced1}
	\begin{aligned}
		&E_t = \\ & A_t  \ {\exp{\biggl(-k_0 \int\displaylimits_{\mathclap{\text{ \tiny along $\bm{\hat{k}_t}$}}} V_I \bm{\hat{k}_a} \cdot (\bm{\hat{k}_t} dr) \biggr)}}\exp{\biggl(jk_0 \int\displaylimits_{\mathclap{\text{\tiny along $\bm{\hat{k}_t}$}}} \ V_R \bm{\hat{k}_t} \cdot (\bm{\hat{k}_t} dr) \biggr)}
	\end{aligned}
\end{equation}
where, $E_t, V_R, V_I$ are functions of $\bm{r}$ and for brevity, this is implicitly assumed in the rest of the paper. In next section, we use ray formulation in (\ref{Eq_raytraced1}) for IPW field inside the extended scatterer for deriving corrections to RA. 

Note that (\ref{Eq_complexwavevectorfield}) is an approximation to the IPW field inside the scatterer as it only includes the first order ray inside the homogeneous scatterer. Due to multiple scattering inside the scatterer's boundaries, there will be higher order rays inside the scatterer, which are ignored in (\ref{Eq_complexwavevectorfield}). Fortunately, for a large, lossy scatterer (considered in this work), it is known that the first order ray is a good approximation \cite{yang2009effective} as higher order rays will contain low energy.

\subsection{Corrections to Conventional RA}
\label{Sec_rayRytov}
To approximate $\nabla \phi_s \cdot \nabla \phi_s$, we start by equating the total field inside the scatterer  (\ref{Eq_rytTfield}) to the ray equation (\ref{Eq_raytraced1}),
\begin{equation}
	\label{Eq_rytHFtotal}
	\begin{aligned}
		& E_i  (\bm{r})  e^{\phi_s(\bm{r})} = \\ & A_t {\exp{\biggl(-k_0 \int V_I(\bm{r}) \cos\theta_t \  dr \biggr)}} \exp{\biggl(jk_0 \int V_R(\bm{r})\  d{r}\biggr) }
	\end{aligned}
\end{equation}
where $\bm{\hat{k}_a} \cdot \bm{\hat{k}_t} = \cos\theta_t$ from Fig. \ref{interface}. Substituting the incident field from (\ref{Eq_complexwavevectorfield}a) as  $E_i(\bm{r}) = A_0(\bm{r})  e^{j k_0 \bm{\hat{k}_i \cdot r}}$ gives,
\begin{equation}
	\label{Eq_rytHFtotal1}
	\begin{aligned}
		\phi_s(\bm{r}) =   \text{ln}\biggl[\frac{A_t}{A_0}\biggr] + k_0 \biggl[j\int V_R\  dr - j\bm{\hat{k}_i \cdot r} - \int V_I \cos\theta_t dr \biggr]
	\end{aligned}
\end{equation}
Note that the quantities $V_R, V_I, E_i$ are functions of $\bm{r}$ in (\ref{Eq_rytHFtotal1}) and for brevity we do not show this dependence in the remainder of this paper. Taking the gradient of (\ref{Eq_rytHFtotal1}) gives (recall from (\ref{Eq_raytraced1}), $\bm{{dr}} =dr \ \bm{\hat{k}_t}$),
\begin{equation}
	\label{Eq_rytHFphasegrad}
	\begin{aligned}
		\nabla \phi_s(\bm{r})=  \biggl[ \nabla    \text{ln}\biggl(\frac{A_t}{A_0}\biggr)\biggr] + k_0\biggl[j  \big(V_R\ \bm{\hat{k}_t} - \bm{\hat{k}_i}\big) - V_I \bm{\hat{k}_a} \biggr]
	\end{aligned}
\end{equation}
so that
\begin{equation}
	\label{Eq_rytHFphasegrad2}
	\begin{aligned}
		&\frac{\nabla  \phi_s  (\bm{r}) \cdot \nabla \phi_s(\bm{r})}{k_0^2} = \\
		& \biggl[V_I^2 - V_R^2 - 1 + 2 V_R (\bm{\hat{k}_t} \cdot \bm{\hat{k}_i})  - 2 j (V_R \bm{\hat{k}_t}-\bm{\hat{k}_i})  \cdot V_I \bm{\hat{k}_a} \biggr] \\ & + \frac{1}{k_0^2} (\nabla \tilde{A} \cdot \nabla \tilde{A} ) 
		 + \frac{2}{k_0} (\nabla \tilde{A}) \biggl[j\big(V_R\ \bm{\hat{k}_t} - \bm{\hat{k}_i}\big) -   V_I \bm{\hat{k}_a} \biggr]
	\end{aligned}
\end{equation}
where, $\tilde{A} = \ln(A_t/A_0)$. We note from Fig. \ref{interface}, $\bm{\hat{k}_i} \cdot \bm{\hat{k}_a} = \cos\theta_i$ and $\bm{\hat{k}_t} \cdot \bm{\hat{k}_i} = \cos\theta_{s}$ where $\theta_{s}$ is the scattering angle. Using this and separating out real and imaginary terms, we can now write (\ref{Eq_rytHFphasegrad2}) as
\begin{equation}
	\label{Eq_rytHFphasegrad3}
	\begin{aligned}
		&\frac{\nabla  \phi_s  (\bm{r}) \cdot \nabla \phi_s(\bm{r})}{k_0^2} = \\
		& \biggl[V_I^2- V_R^2 - 1 + 2 V_R \cos\theta_{s}  + \frac{\nabla \tilde{A} \cdot \nabla \tilde{A}}{k_0^2} -  \frac{2(\nabla \tilde{A})  V_I \bm{\hat{k}_a}}{k_0} \biggr]  \\
		& + 2j \biggl[ (V_I \cos\theta_i - V_R V_I \cos\theta_t ) + \frac{1}{k_0} (\nabla \tilde{A})\big(V_R\ \bm{\hat{k}_t} - \bm{\hat{k}_i}\big) \biggr].
	\end{aligned}
\end{equation}
Equation (\ref{Eq_rytHFphasegrad3}) provides an expression for $\nabla \phi_s \cdot \nabla \phi_s$ which is required in RI (\ref{Eq_rytov2}b) (neglected in RA (\ref{Eq_RA}b)). Expanding the contrast function (\ref{Eq_rytov2}b) of RI using (\ref{Eq_effective}) gives,
\begin{equation}
	\label{Eq_rytovfulldB4}
	\begin{aligned}
		\chi_{\text{RI}}(\bm{r}) &= (\nu_R+j \nu_I)^2-1 +\frac{\nabla  \phi_s  (\bm{r}) \cdot \nabla \phi_s(\bm{r})}{k_0^2} \\
		& = V_R^2 - V_I^2 + 2j V_R V_I \cos\theta_t + \frac{\nabla  \phi_s  (\bm{r}) \cdot \nabla \phi_s(\bm{r})}{k_0^2}
	\end{aligned}
\end{equation}

Substituting $(\nabla \phi_s \cdot \nabla \phi_s)/k_0^2$ from (\ref{Eq_rytHFphasegrad3}) to (\ref{Eq_rytovfulldB4}) leads to cancellation of several terms and gives,
\begin{equation}
	\label{Eq_rytovfulldB41}
	\begin{aligned}
		\chi_{\text{RI}}&(\bm{r}) = {\bigg[2 V_R \cos\theta_{s} -2 + \frac{1}{k_0^2} (\nabla \tilde{A} \cdot \nabla \tilde{A} ) - \frac{2}{k_0} (\nabla \tilde{A}) V_I \bm{\hat{k}_a}  \bigg]} \\
		& \qquad \qquad + j {\bigg[ 2 V_I \cos\theta_i + \frac{2}{k_0} (\nabla \tilde{A})\big(V_R(\bm{r})\ \bm{\hat{k}_t} - \bm{\hat{k}_i}\big) \bigg]}.
	\end{aligned}
\end{equation}
Equation (\ref{Eq_rytovfulldB41}) can be further modified using (\ref{Eq_VRVIlowloss1}) and (\ref{Eq_VRVIlowloss2}) to replace $V_R$ and $V_I$ in terms of $\nu_R$ and $\nu_I$ under low-loss conditions as,
\begin{equation}
	\label{Eq_rytovfulldB5}
	\begin{aligned}
		\chi_{\text{RI}}&(\bm{r}) = {\bigg[2 (\nu_R \cos\theta_{s} -1) + \frac{1}{k_0^2} (\nabla \tilde{A} \cdot \nabla \tilde{A} ) - \frac{2}{k_0} (\nabla \tilde{A}) V_I \bm{\hat{k}_a}  \bigg]}  \\
		& + j {\bigg[  2 \frac{\nu_R \nu_I}{\sqrt{\nu_R^2 - \sin^2\theta_i}} \cos\theta_i + \frac{2}{k_0} (\nabla \tilde{A})\big(\nu_R(\bm{r})\ \bm{\hat{k}_t} - \bm{\hat{k}_i}\big) \bigg] }
	\end{aligned}
\end{equation}
Complex refractive index can also be expressed in terms of complex permittivity using (\ref{Eq_complexNu_R_LL}). Using this, the final expression for the contrast function in RI under low-loss, high frequency conditions is given by
\begin{equation}
	\label{Eq_rytovfulldB6}
	\begin{aligned}
		\chi_{\text{RI}}&(\bm{r}) = \\ &\biggl(\underbrace{2 (\sqrt{\epsilon_R} \cos\theta_{s} -1)}_{\text{R}_1} + \underbrace{\frac{1}{k_0^2} (\nabla \tilde{A} \cdot \nabla \tilde{A} )}_{\text{R}_2 \text{ (crosstalk)}} - \underbrace{\frac{2}{k_0} (\nabla \tilde{A}) V_I \bm{\hat{k}_a}}_{\text{R}_3 \text{ (crosstalk)}}  \biggr)\\ 
		&+ j {\biggl( \underbrace{\frac{\epsilon_I}{\sqrt{\epsilon_R - \sin^2\theta_i}} \cos\theta_i}_{\text{I}_1} + \underbrace{\frac{2}{k_0} (\nabla \tilde{A})\big(\nu_R(\bm{r})\ \bm{\hat{k}_t} - \bm{\hat{k}_i}}_{\text{I}_2 \text{ (crosstalk)}}\big) \biggr) }
	\end{aligned} 
\end{equation}

 Unlike contrast function ${\chi_{\text{RA}}}$ of conventional RA, the derived, corrected contrast ${\chi_{\text{RI}}}$ is a non-linear function of permittivity. Furthermore, the imaginary part $\operatorname*{Im}({\chi_{\text{RI}}})$ depends on both the real and imaginary parts of the permittivity and this describes the ``crosstalk" where even when the permittivity is real, there will be a component in the imaginary part of the contrast function. Similarly, the real part of the contrast function also depends on both the real and imaginary parts of the permittivity.

Further simplification of (\ref{Eq_rytovfulldB6}) is possible. For the high frequency regime where $k_0$ is large, we can approximate (\ref{Eq_rytovfulldB6}) by ignoring the cross terms (R$_2$, R$_3$, I$_2$). This approximation will be valid as long as the spatial variation of the term $\tilde{A}=\ln (A_t/A_0)$ is small (so that $\nabla \tilde{A}$ is small). Even for moderately high frequencies, the cross terms will be smaller due to division by $k_0$ terms. Furthermore, for large homogeneous scatterers, the gradient of $\tilde{A}$ will be minimal inside and outside the objects. On the boundaries there will be a discontinuity and hence our approximations will generally be accurate everywhere except at the boundaries of the objects where we can expect some errors. Based on these approximations, we can ignore the cross terms and rewrite (\ref{Eq_rytovfulldB6}) as,
\begin{equation}
	\label{Eq_rytovfulldB6-1}
	\begin{aligned}	
		\chi_{\text{RI}}(\bm{r}) &=2 (\sqrt{\epsilon_R} \cos\theta_{s} -1)
		+ j  \frac{\epsilon_I}{\sqrt{\epsilon_R - \sin^2\theta_i}} \cos\theta_i	
	\end{aligned} 
\end{equation}
Using Fig. \ref{interface}, (\ref{Eq_SnellLaw1}) and (\ref{Eq_VRVIlowloss1}), we can write the scattering angle as,
\begin{equation}
	\label{Eq_costhetas}
	\begin{aligned}	
		\cos\theta_s &= \cos(\theta_i-\theta_t)\\
		& = \cos\theta_i \cos\theta_t + \sin\theta_i \sin\theta_t\\
		& = \frac{\cos\theta_i \sqrt{\nu_R^2 - \sin^2 \theta_i}}{\nu_R} + \frac{\sin^2\theta_i}{\nu_R}
	\end{aligned} 
\end{equation}
Substituting (\ref{Eq_costhetas}) back in (\ref{Eq_rytovfulldB6-1}) gives the final expression for $\chi_{\text{RI}}$ as
\begin{equation}
	\label{Eq_rytovfulldB6_2}
	\begin{aligned}	
		\chi_{\text{RI}} &=2 \cos\theta_i( \sqrt{\epsilon_R - \sin^2 \theta_i} - \cos\theta_i)
		+ j  \frac{\epsilon_I \cos\theta_i}{\sqrt{\epsilon_R - \sin^2\theta_i}}  	
	\end{aligned} 
\end{equation}
This is our proposed contrast function which is also valid under strong scattering as the the term $\nabla \phi_s \cdot \nabla \phi_s$ is not neglected unlike in RA. 

In the derivation of $\chi_{\text{RI}}$, we do not impose any restriction on its permittivity value. We only impose a low-loss condition ($\epsilon_R \gg \epsilon_I$ so that $|\epsilon_r|$ can be arbitrarily large). As a result, unlike RA, the contrast function $\chi_{\text{RI}}$ in (\ref{Eq_rytovfulldB6_2}) becomes a non-linear function of permittivity and this provides a fundamentally new extension to RA which is valid for even strongly scattering objects that have a small loss tangent.

We can also look at (\ref{Eq_rytovfulldB6_2}) from the perspective of Fermat's principle to gain more insight. Under strong scattering ($\epsilon_R\gg1$), and for the special case of normal incidence ($\theta_i=0$), our result  (\ref{Eq_rytovfulldB6_2}) reduces to refractive index $\chi_{\text{RI}} \approx 2 (\nu_R-1) + 2j \nu_I = 2(\nu-1)$. This agrees with Fermat's principle where the incremental phase change of a ray is directly related to the product of the path length along the ray and refractive index contrast $(\nu(\bm{r}) -1)$. In other words, the incremental phase change of a ray per wavelength should be proportional to $k_0(\nu-1)$. For conventional RA, it is known (using asymptotic techniques) that the incremental phase change per wavelength is $\frac{1}{2} k_0 (\nu^2 -1)$ which does not match the expected phase change as per Fermat's principal (\ref{Eq_rytovfulldB6_2}). Therefore, xRA-LM also appears to better satisfy the underlying physics of the problem. 

\subsection{Simplification}
To further simplify the contrast function, $\chi_{\text{RI}}$ in (\ref{Eq_rytovfulldB6_2}), we remove its dependence on $\theta_i$ without significantly compromising accuracy. Estimation of $\theta_i$ is plausible in direct problems as the information about scatterer's shape is known. But it will require intricate geometric calculations which defeats one of the purposes of using xRA, i.e., a computationally straightforward, non-iterative linear alternative to VSI. 

In a typical scattering setup, the incident rays enter the scattering object from a wide direction of incidence directions in the range $\theta_i \in [-\pi/2, \pi/2]$. Therefore, we can remove the dependence on $\theta_i$ by averaging ${\chi_{\text{RI}}}$ uniformly over this range of $\theta_i$ to obtain
\begin{equation}
	\label{Eq_avgIm}
	\begin{aligned}
		{{\tilde{\chi}_{\text{RI}}}} &= \frac{1}{\pi} \int_{-\frac{\pi}{2}}^{\frac{\pi}{2}} \chi_{\text{RI}} \ d\theta_i \\
		& = \frac{1}{\pi} \int_{-\frac{\pi}{2}}^{\frac{\pi}{2}} \bigg(2 \cos\theta_i( \sqrt{\epsilon_R - \sin^2 \theta_i} - \cos\theta_i)
		\ + \\ & \qquad \qquad \qquad \qquad \qquad \qquad  j  \frac{\epsilon_I \cos\theta_i}{\sqrt{\epsilon_R - \sin^2\theta_i}} \bigg) \ d\theta_i
	\end{aligned}
\end{equation}
This integral can be solved analytically (using substitution $u = \sin\theta_i$) to obtain
\begin{equation}
	\label{Eq_avgIm1}
	\begin{aligned}
		{{\tilde{\chi}_{\text{RI}}}}  = \frac{2}{\pi}\biggl( \sqrt{\epsilon_R-1} + \sin^{-1}\biggl[\frac{1}{\sqrt{\epsilon_R}}\biggr] - \frac{\pi}{2} \biggr) \\ + \  j\frac{2 }{\pi} \epsilon_I \sin^{-1} \biggl[\frac{1}{\sqrt{\epsilon_R}}\biggr]
	\end{aligned}
\end{equation}  
It should be noted that averaging over the range of incident angles will lead to errors and can be considered as localizing the effect of the incident angles \cite{chen2018computational, habashy1993beyond, murch1992extended}. However, we show in the simulation that this error is acceptable ($< 15$\%) even under strong scattering conditions. Also, even though ${{\tilde{\chi}_{\text{RI}}}}$ is derived using ray approximation inside the lossy media, it is also correct for background vacuum/air where ${{\tilde{\chi}_{\text{RI}}}} = 0 +0j$ since $\epsilon_R=1$ and $\epsilon_I = 0$. Therefore, ${{\tilde{\chi}_{\text{RI}}}}$ can be used as a physical parameter to characterize the permittivity distribution inside the DoI.

Substituting ${{\tilde{\chi}_{\text{RI}}}}$ as $\chi_{\text{RI}}$ in (\ref{Eq_rytov2}) gives the proposed xRA-LM approximation which can be written as,
\begin{equation} 
	\label{Eq_RIfinal2}
	\begin{aligned}
		E(\bm{r})  &= E_i(\bm{r})  \exp\bigg (\int_{\mathcal{D}} \mathcal{H}(\bm{r}, \bm{r'}) {{\tilde{\chi}_{\text{RI}}}}(\bm{r}') d\bm{r'}^2 \bigg),
	\end{aligned} 
\end{equation} 
where $\mathcal{H}(\bm{r}, \bm{r'})$ is denoted here as the sensitivity kernel,
\begin{equation}
	\label{Eq_RI_SVK}
	\begin{aligned}
		\mathcal{H}(\bm{r}, \bm{r'})  =  \frac{k_0^2}{E_i(\bm{r})} g(\bm{r}, \bm{r'})  E_i(\bm{r'}),  \quad \bm{r}' \in \mathcal{D}, \bm{r} \in \mathcal{B}
	\end{aligned} 
\end{equation}
The total field in (\ref{Eq_RIfinal2}) can also be  decomposed in terms of attenuation and phase change components as,
\begin{equation}
	\label{Eq_RIfinal3}
	\begin{aligned}
		E(\bm{r})  = E_i(\bm{r}) \cdot & \exp\biggl({ \int_{\mathcal{D}} \bigl[  \mathcal{H}^R \tilde{\chi}_{\text{RI}}^R - \mathcal{H}^I \tilde{\chi}_{\text{RI}}^I  \bigr]d\bm{r'}^2}\biggr) \cdot \\ & \ \ \ \exp\biggl({j\int_{\mathcal{D}}  \bigl[ \mathcal{H}^R \tilde{\chi}_{\text{RI}}^I + \mathcal{H}^I \tilde{\chi}_{\text{RI}}^R \bigr]d\bm{r'}^2 }\biggr)
	\end{aligned} 
\end{equation}
where, $\mathcal{H}^R(\bm{r}, \bm{r}')$ and $\mathcal{H}^I(\bm{r}, \bm{r}')$ are real and imaginary parts of the sensitivity kernel $\mathcal{H}(\bm{r}, \bm{r}')$ whereas $\tilde{\chi}_{\text{RI}}^R(\bm{r}')$ and $\tilde{\chi}_{\text{RI}}^I(\bm{r}')$ are real and imaginary part of contrast $\tilde{\chi}_{\text{RI}}(\bm{r}')$. 

\subsection{Modifications for Extremely Strong Scattering}
\label{Sec_final_form}

As we show later, our xRA-LM formulation in (\ref{Eq_RIfinal3}) is accurate even under strong scattering (large scatterers with $2 \le \epsilon_R < 5$) and provides error less than 15\% in predicting wave scattering. Hence xRA-LM already surpasses existing non-iterative linear scattering models. However, for extremely strong scattering (large scatterers with $5 \le \epsilon_R \le 50$), the errors increase (20\% to 30\%), but it is still significantly better than any other existing non-iterative linear methods \cite{chen2018computational, chen2020review}. In this section, we suggest a minor modification to (\ref{Eq_RIfinal3}) to increase its accuracy for extremely strong scattering conditions.

In (\ref{Eq_rytovfulldB6-1}), the real part of the contrast ${\chi}_{\text{RI}}^R$ depends on the scattering angle $\theta_s$ (which further depends on incident angle $\theta_i$, and the scatterer's permittivity distribution) and we approximate it in (\ref{Eq_costhetas}) as its exact value is difficult to estimate. On the otherhand, the imaginary part of contrast ${\chi}_{\text{RI}}^I$ only depends on $\theta_i$ and therefore the real component will likely be significantly less accurate than the imaginary part of the contrast function. This difference in accuracy will be severe under extremely strong scattering conditions where our approximation to $\theta_s$ can be less accurate and the error in the real part of contrast ${\chi}_{\text{RI}}^R$ corresponding large. 

By considering the line of sight (LOS) path or region between the transmitter and receiver, we can attempt to reduce the possible error in the real part of the contrast function ${\chi}_{\text{RI}}^R$ by realizing the LOS region will be dominated by attenuation rather than more intricate scattering effects \cite{yang2009effective}. We can therefore take a straightforward step and neglect ${\chi}_{\text{RI}}^R$ within the LOS region and only rely on ${\chi}_{\text{RI}}^I$ for estimating the total field at the receiver. Intuitively, this means that for points within the LOS region, we rely on the absorption caused by the scatterer rather than estimating the effect of higher order scattering (inside scatterer). Even with small loss tangent $\epsilon_R \gg \epsilon_I$, since $\epsilon_R$ is large, the value of $\epsilon_I$ can be large enough to cause substantial absorption of electromagnetic energy (especially for scatterers large in size). This is also supported by numerical studies \cite{yang2009effective} which show that for scatterers with substantial loss, the higher-order scattered rays inside the scatterer are weak. Hence, the total field within LOS regions will be dominated by the attenuating incident field. For points outside the LOS region the incident field will not dominate and we cannot rely on absorption alone and hence cannot ignore $\chi^R_{\text{RI}}$.  

To define the LOS region in the scatterer, we use Fresnel zones. We use $\mathcal{E}$ to denote the set of points lying inside the first Fresnel zone for a source and receiver pair. Using our approach, we substitute ${\chi}_{\text{RI}}^R = 0$ in (\ref{Eq_RIfinal3}) for points that lie inside the first Fresnel zone $\mathcal{E}$ so that (\ref{Eq_RIfinal3}) becomes,
\begin{equation}
	\label{Eq_RIfinal5}
	\begin{aligned}
		E (\bm{r})   =  E_i(\bm{r}) & \exp  \biggl( \int_{\mathcal{D}} \biggl[ \mathcal{H}^R (\beta \tilde{\chi}_{\text{RI}}^R) - \mathcal{H}^I \tilde{\chi}_{\text{RI}}^I \biggr]d\bm{r'}^2 \biggr) \\ &\exp\biggl(j \int_{\mathcal{D}} \biggl[ \mathcal{H}^R \tilde{\chi}_{\text{RI}}^I + \mathcal{H}^I (\beta \tilde{\chi}_{\text{RI}}^R) \biggr]d\bm{r'}^2 \biggr),
	\end{aligned}
\end{equation}
where,
\begin{equation}
	\label{Eq_RIfinal6}
	\begin{aligned}
		\beta(\bm{r}, \bm{r}') = 
		\begin{cases}
			0 &  \text{if} \ \ \bm{r}' \in \mathcal{E} \text{  and  } \epsilon_R \gg 1\\
			1 & \text{otherwise},
		\end{cases}
	\end{aligned}
\end{equation}
By using (\ref{Eq_RIfinal6}), the real part of the contrast is set to zero for points of the scatterer in the first Fresnel zone when calculating the field. For all other points, (\ref{Eq_RIfinal5}) reduces exactly to (\ref{Eq_RIfinal3}).

While the correction (\ref{Eq_RIfinal5}) is somewhat heuristic in justification, we show in the next section it performs extremely well for extremely strong scattering conditions (for large scatterers with $\epsilon_r>5$), when compared to the results without the correction (\ref{Eq_RIfinal3}). It is also important to note that even without the LOS corrections in (\ref{Eq_RIfinal5}), our derived xRA-LM method performs accurately for $\epsilon_r\le 5$ (as shown later) and for all scattering angles, which already significantly surpasses any other non-iterative linear methods. Future work can focus further on improved approximations to $\theta_s$ to enhance the technique further.

\section{Numerical Results}
\label{Sec_results}
This section presents numerical results for investigating the performance of xRA-LM for solving direct problems. We compare the results obtained using xRA-LM with three techniques, 1) Numerical solution using MoM in VSI, 2) RA, and 3) BA. All the results shown are for (\ref{Eq_RIfinal5}) while comparisons with (\ref{Eq_RIfinal3}) are also provided later to show the accuracy of the underlying technique.  

We quantify the accuracy of the techniques by defining the relative error between the magnitude of the exact total field $E_\text{ex}$ (using MoM) and the estimated total field $\tilde{E}$ (using xRA-LM, BA or RA) as,
\begin{equation}
\label{Eq_RE}
\text{RE } (\text{in } \%) = \bigg|\frac{|E_\text{ex}|-|\tilde{E}|}{|E_\text{ex}|}\bigg| \times 100
\end{equation}

\begin{figure}[!h]
	\centering
	\includegraphics[width=3.2in]{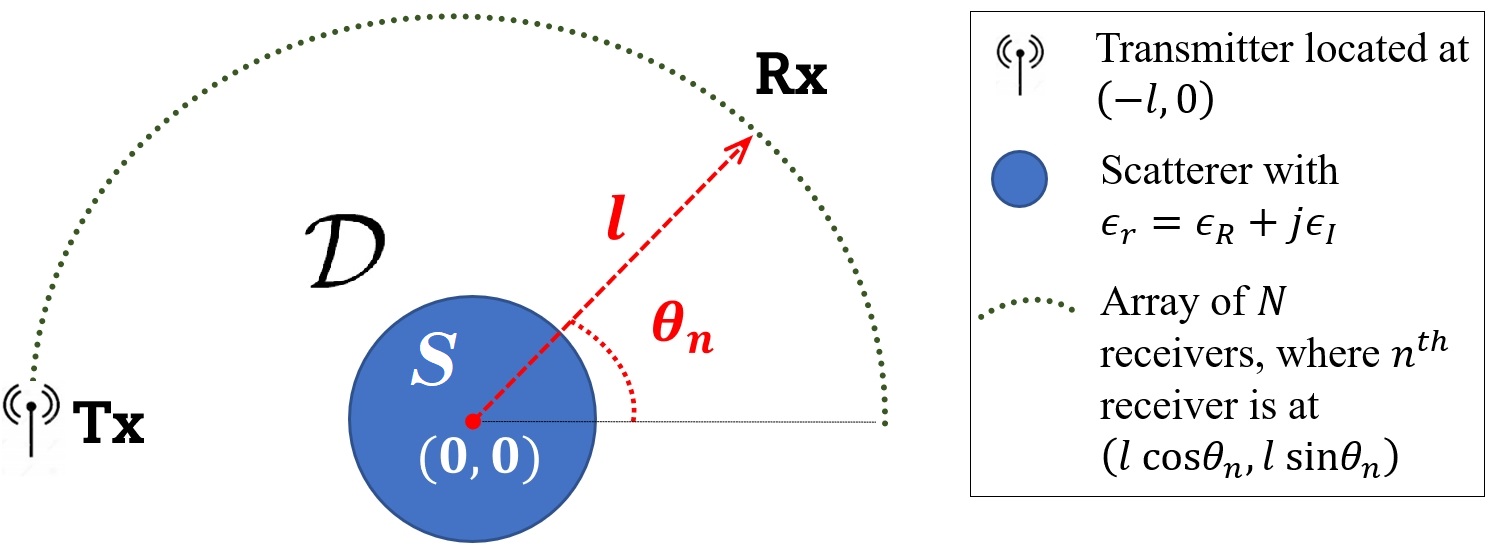}
	\caption{Illustration of wave scattering setup for a circular scattering object $\mathcal{{S}}$ centered at origin as described in the text. Note that figure is scaled for better visualization and does not represent actual actual distances between Tx, Rx and the scatterer. }
	\label{NumSetup}
	\vspace{-0.2\baselineskip}
\end{figure}

\subsection{Simulation Setup}
\label{Sec_setup}
\begin{figure}[!h]
	\centering
	\includegraphics[width=3in]{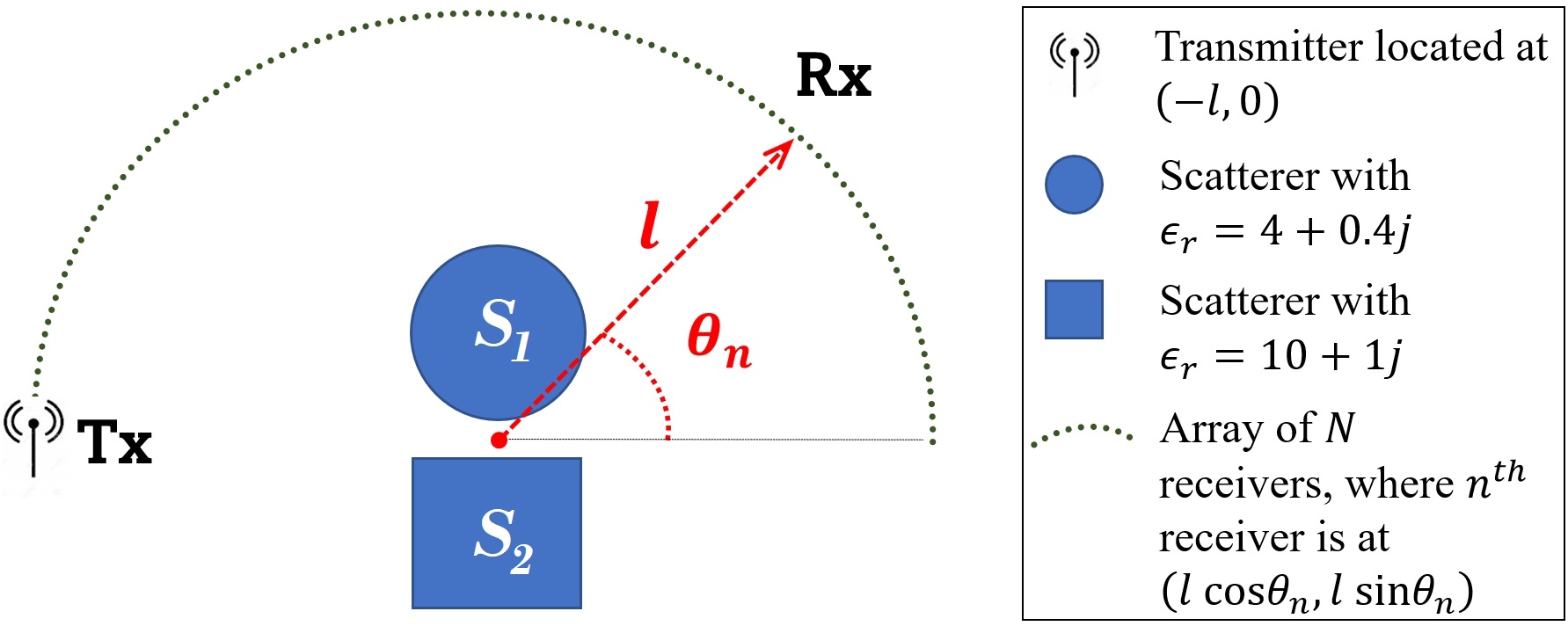}
	\caption{Illustration of wave scattering setup for a circular and square cylinder as describe din the text. Location of receivers, transmitters and origin are all same as Fig. \ref{NumSetup}. Note that figure is scaled for better visualization and do not represent actual actual distances between Tx, Rx and the scatterer.}
	\label{NumSetup2}
	\vspace{-0.2\baselineskip}
\end{figure}

\begin{figure}[!h]
	\centering
	\includegraphics[width=3in]{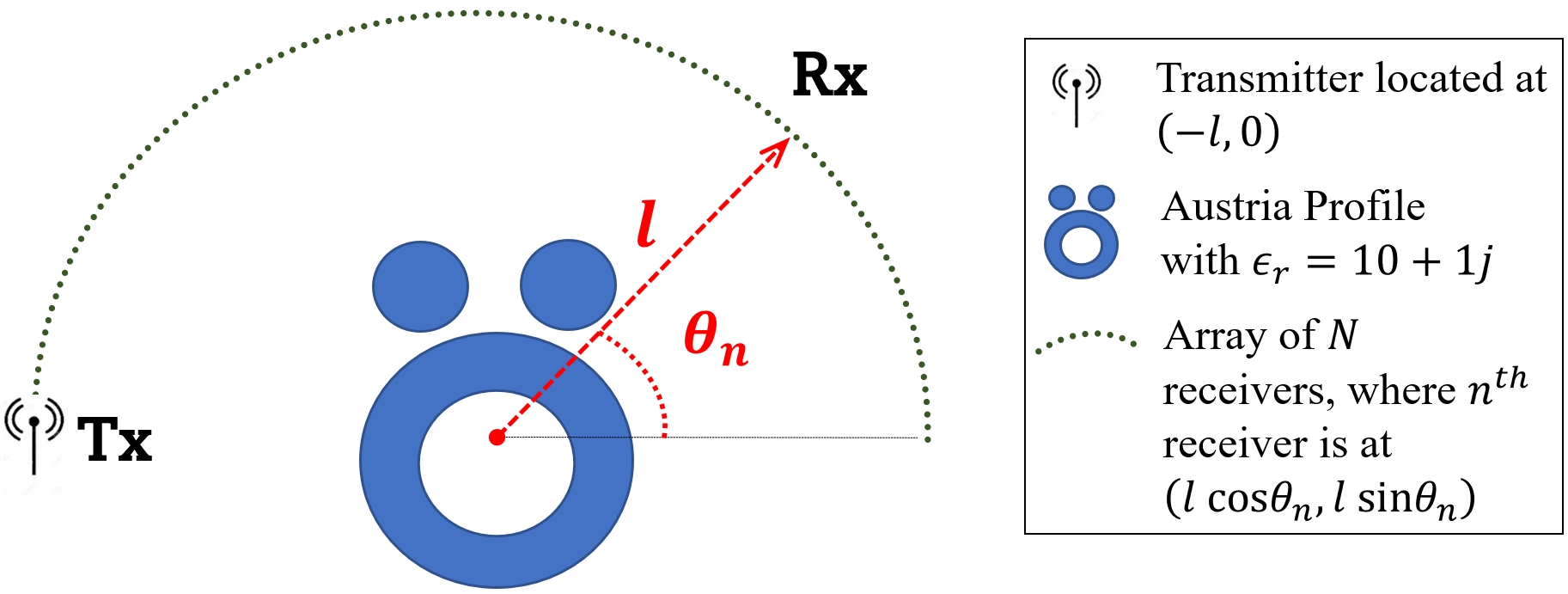}
	\caption{Illustration of wave scattering setup for the Austria profile. Location of receivers, transmitters and origin are all same as in Fig. \ref{NumSetup} and Fig. \ref{NumSetup2}. The size of the two disks and the ring in the Austria profile is defined in terms of the incident wavelength. The diameter of disks is $0.54 \times \lambda_0$. The ring has exterior diameter of $1.6 \times \lambda_0$ and inner diameter is $0.8 \times \lambda_0$. The distance between centers of two disks is $0.8 \times \lambda_0$ and vertical distance between center of discs and ring is $\lambda_0$. In our simulations frequency is taken as 2.4 GHz ($\lambda_0 = 12.5$ cm). }
	\label{NumSetup3}
	\vspace{-0.2\baselineskip}
\end{figure}

The simulation setup for the direct problem follows that shown in Fig. \ref{geometry} in which the frequency of the electromagnetic radiation is taken as 2.4 GHz ($\lambda_0 = 12.5$ cm). We consider three scatterer profiles as shown in Fig. \ref{NumSetup}, Fig. \ref{NumSetup2} and Fig. \ref{NumSetup3}. The scatterer shown in Fig. \ref{NumSetup} is centered at $(0, 0)$ and is a circular cylinder with diameter $10 \times \lambda_0 = 1.25$ m and constant $\epsilon_r = \epsilon_R + j \epsilon_I$. The circular cylinder can be considered a standard test object in wave scattering evaluation \cite{chew1999waves, Mittra1998, gibson2021method, enright1992towards, pranay2020}. The circular cylinder is illuminated by a source (\textbf{Tx}) of monochromatic, time harmonic, vertically polarized electromagnetic radiation at \textbf{Tx} is $(-3, 0)$ m. An array of $N$ receivers are placed along the semicircular boundary $\mathcal{B}$ with radius $l = 24 \times \lambda_0 = 3$ m around the scatterer. The location of the $n^{th}$ receiver is $(3 \cos\theta_n, 3 \sin\theta_n)$ m. 

The second scatterer profile has two cylinders as shown in Fig. \ref{NumSetup2}, and is exactly the same as Fig. \ref{NumSetup} in terms of location of source, receivers and size of DoI. Scatterer $\mathcal{S}_1$ is a circular cylinder centered at $(0, 0.44)$ m and $\mathcal{S}_2$ is a square cylinder centered at $(0, -0.44)$ m (both with infinite height along z-axis). The diameter of the circular cylinder and the side of the square cylinder are both $5\times \lambda_0 = 0.625$ m.  

The third scatterer profile in Fig. \ref{NumSetup3} is the well-known Austria profile which is often used as a benchmark profile in inverse scattering literature \cite{chen2018computational, chen2020review, Xudongchen}. The details of the Austria profile are provided in the caption of Fig. \ref{NumSetup3} (note that the size and location of the two disks in the Austria profile are expressed in terms of the incident wavelength).

For generating results, we vary the real part of the relative permittivity of the scatterers between $1.1 < \epsilon_R \le 50$ and $\delta = \epsilon_I/\epsilon_R \in [10^{-4}, 10^{-1}]$ (the complex valued relative permittivity can be written as $\epsilon_r = \epsilon_R(1+\delta j)$). We select this range of complex-valued permittivity for numerical tests based on real world objects \cite{4562803, Productnote, ahmad2014partially, Dubey2021}. Values of $\epsilon_R$ for objects in the environment around us vary from $2< \epsilon_R \le 50$ where $\epsilon_R \ge 20$ is for water and human body at 2.4 GHz, at room temperature \cite{4562803, Productnote, ahmad2014partially, Dubey2021}. Therefore, we used this range of $\epsilon_R$ in our numerical tests. The loss tangent ($\delta = \epsilon_I/\epsilon_R \in [10^{-4}, 10^{-1}]$) considered also represents realistic values for objects around us at 2.4 GHz \cite{4562803, Productnote, ahmad2014partially, Dubey2021}.

For numerical simulation, we divide the DoI (in both Fig. \ref{NumSetup} and Fig. \ref{NumSetup2}) into $M$ small grids, each of size $\lambda/10$ where $\lambda = \lambda_0/\sqrt{|\epsilon_r|}$. The grid size of $\lambda/10$ is an accepted convention to ensure sufficient accuracy \cite{chen2018computational}. 

	\begin{figure*}[!h]
	\captionsetup[subfigure]{aboveskip=2pt,belowskip=5pt, font=footnotesize,oneside,margin={0.4cm,0cm}}
	\centering
	\begin{subfigure}{0.42\textwidth}
		\includegraphics[width=\textwidth]{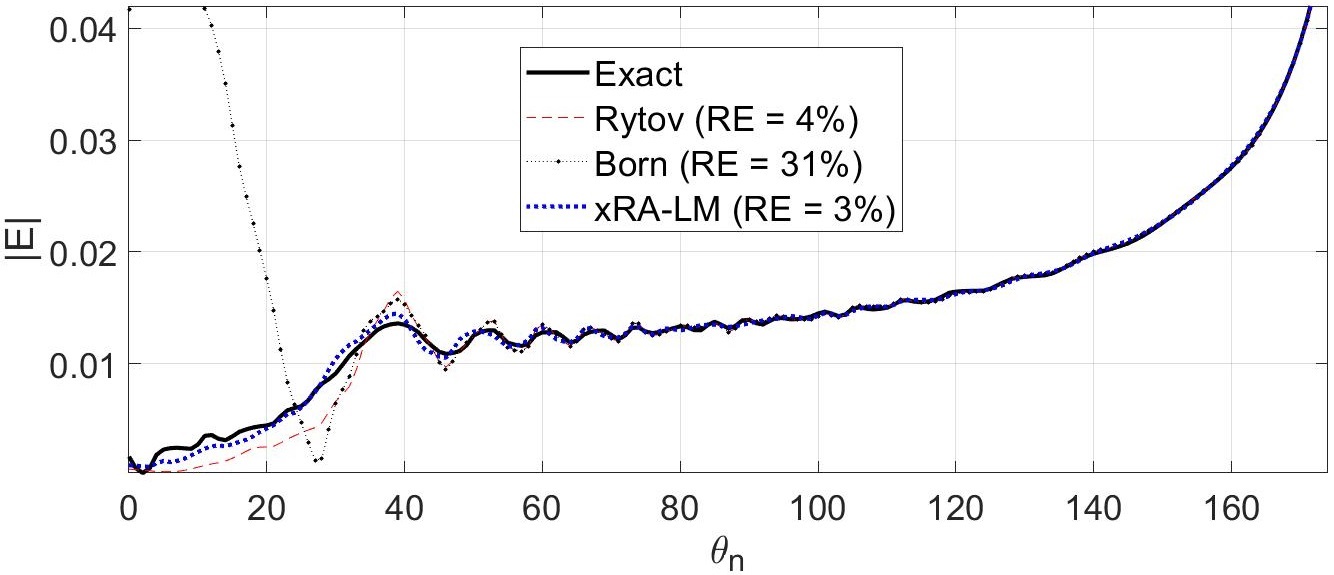}
		\subcaption{$\bm{\epsilon_r = 1.1+0.11j}$}
	\end{subfigure}       
	\begin{subfigure}{0.42\textwidth}
		\includegraphics[width=\textwidth]{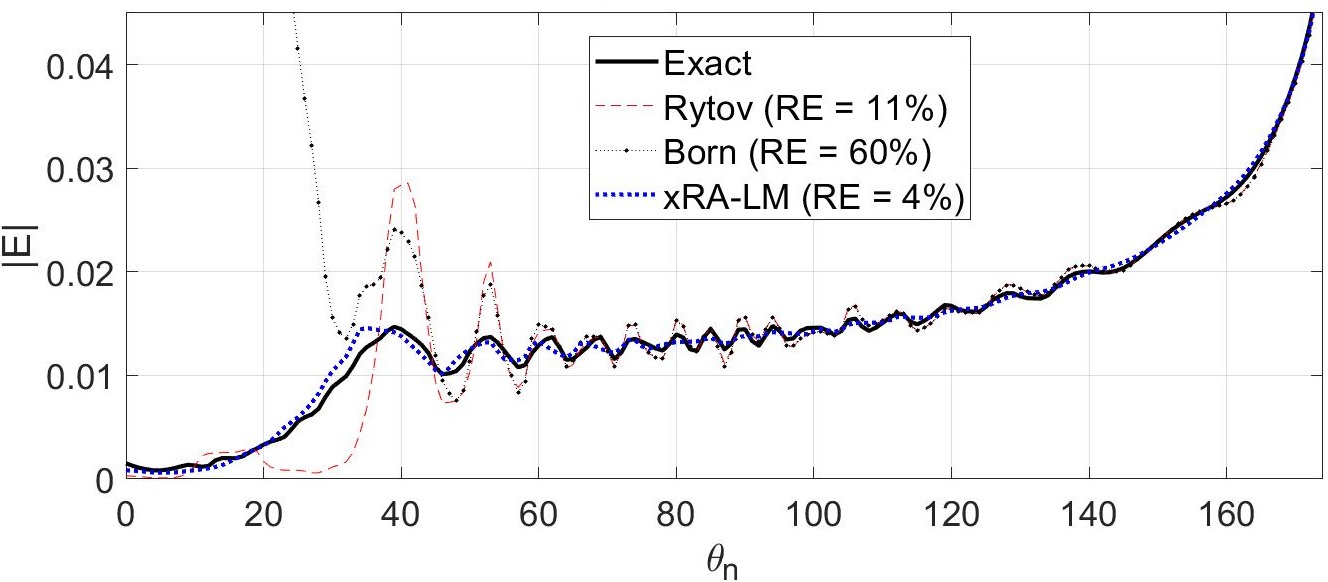}
		\subcaption{$\bm{\epsilon_r = 1.5+0.15j}$}
	\end{subfigure}	
	\begin{subfigure}{0.42\textwidth}
		\includegraphics[width=\textwidth]{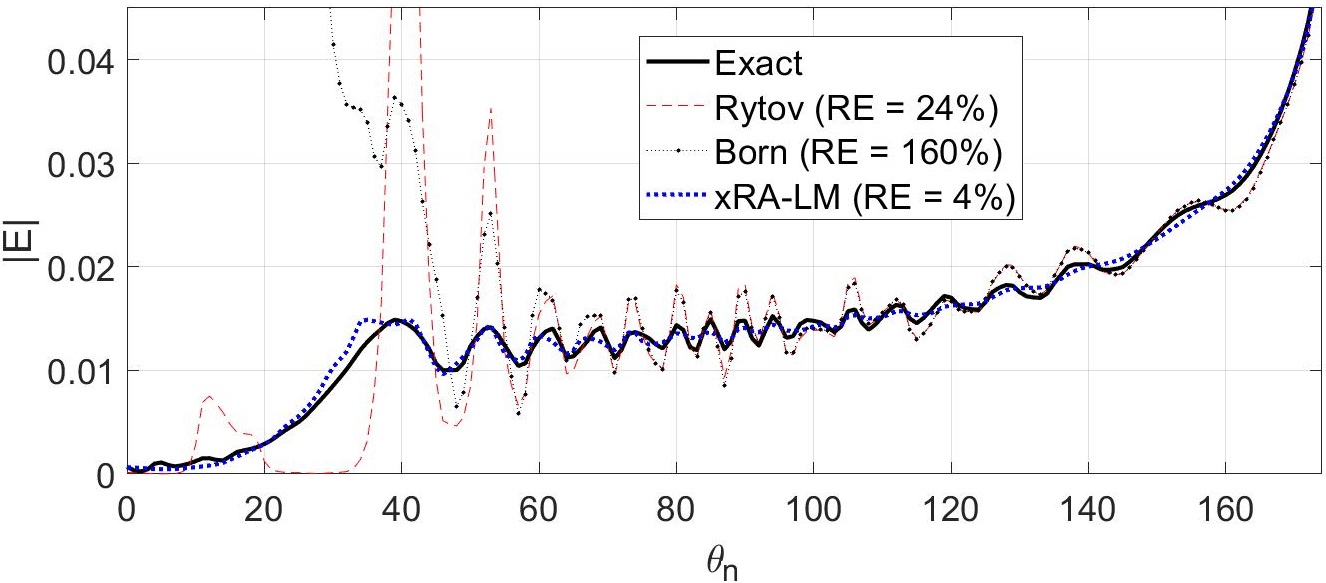}
		\subcaption{$\bm{\epsilon_r = 2+0.2j}$}
	\end{subfigure}       
	\begin{subfigure}{0.42\textwidth}
		\includegraphics[width=\textwidth]{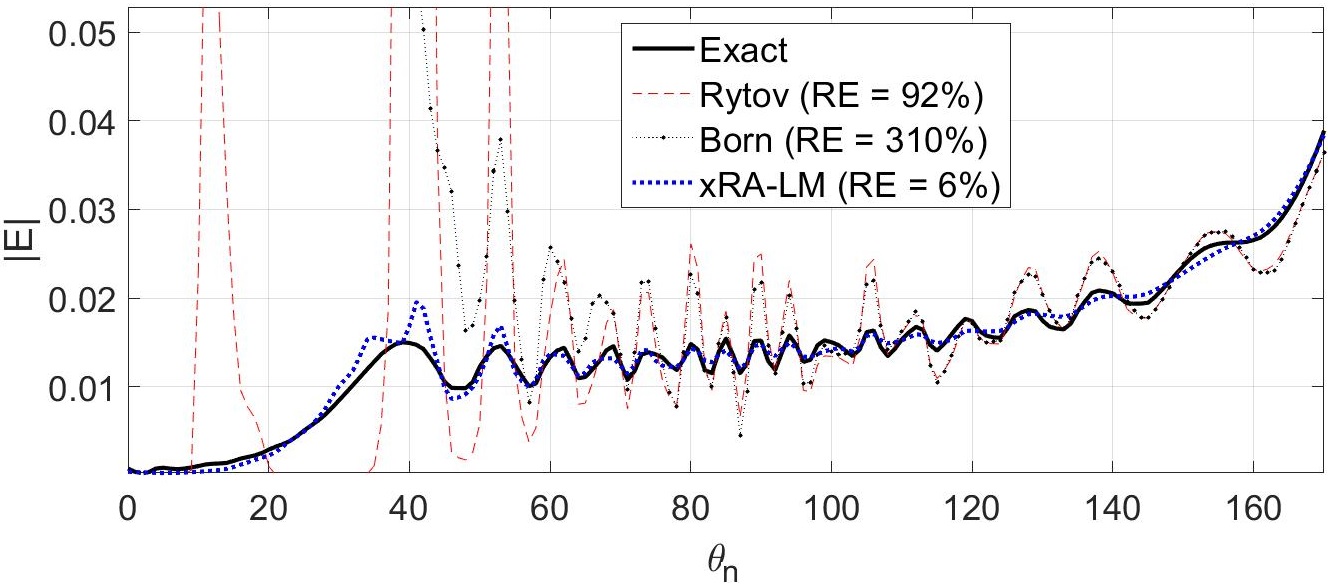}
		\subcaption{$\bm{\epsilon_r = 3+0.3j}$}
	\end{subfigure}	
	\begin{subfigure}{0.42\textwidth}
		\includegraphics[width=\textwidth]{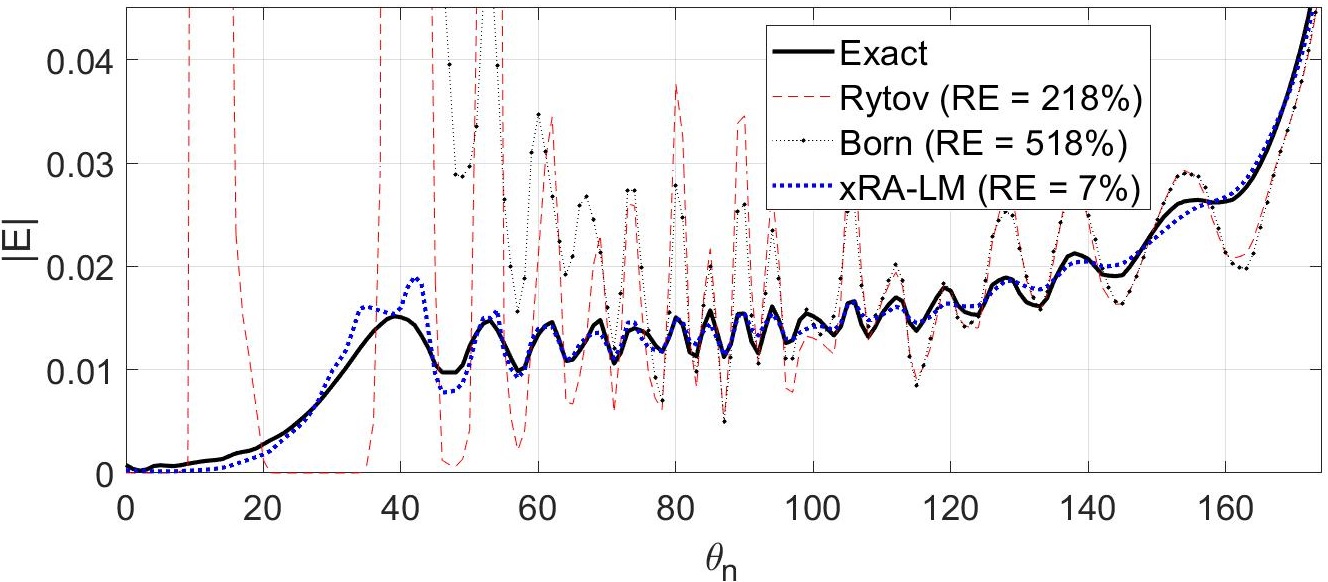}
		\subcaption{$\bm{\epsilon_r = 4+0.4j}$}
	\end{subfigure}       
	\begin{subfigure}{0.42\textwidth}
		\includegraphics[width=\textwidth]{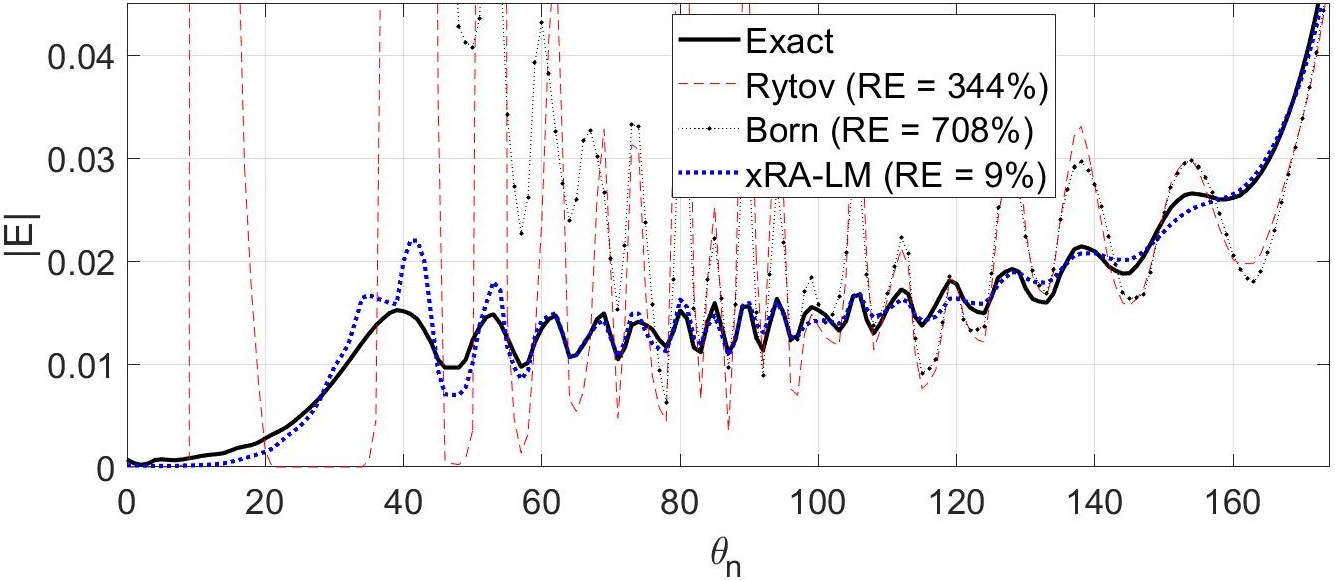}
		\subcaption{$\bm{\epsilon_r = 5+0.5j}$}
	\end{subfigure}	
	\begin{subfigure}{0.42\textwidth}
		\includegraphics[width=\textwidth]{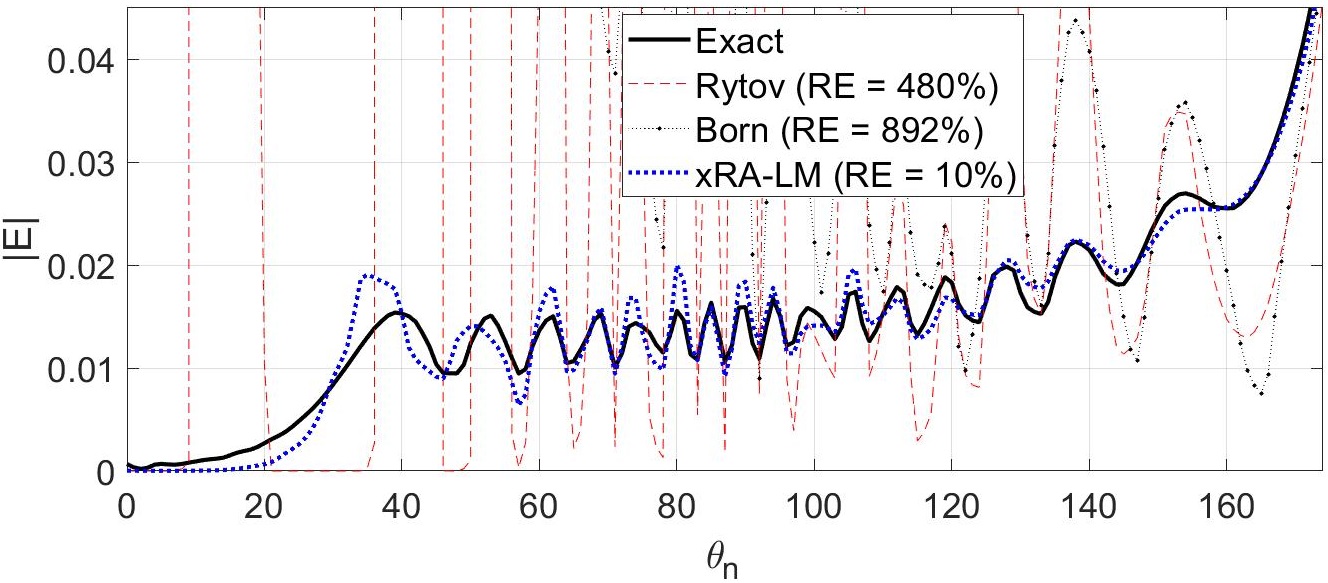}
		\subcaption{$\bm{\epsilon_r = 10+1j}$}
	\end{subfigure}       
	\begin{subfigure}{0.42\textwidth}
		\includegraphics[width=\textwidth]{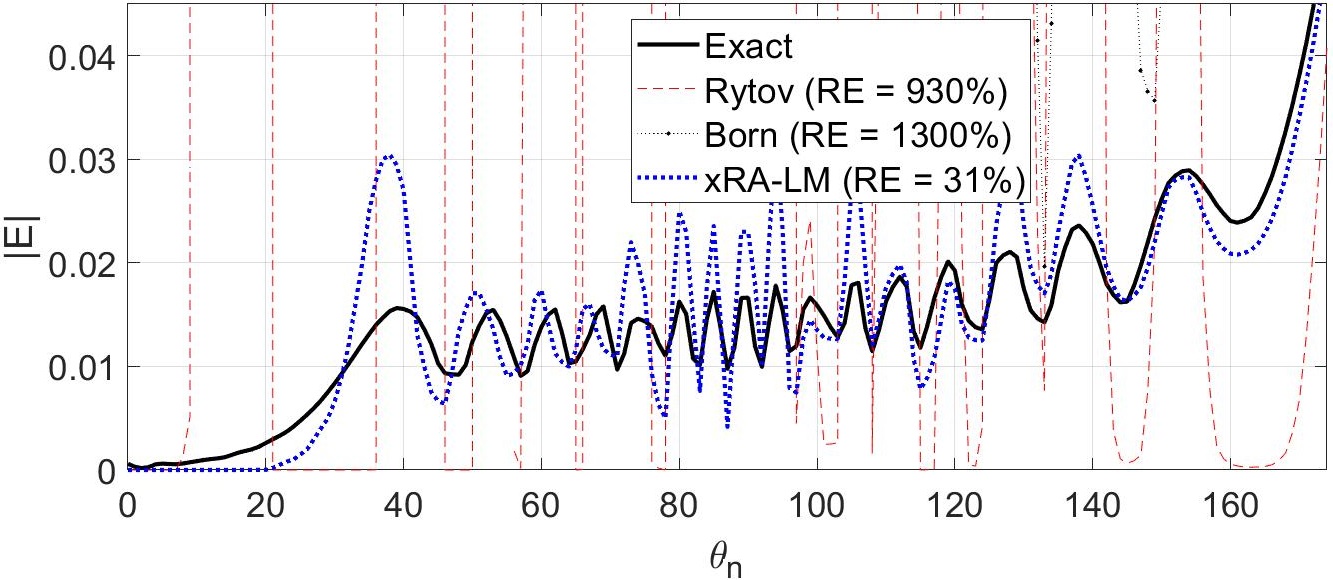}
		\subcaption{$\bm{\epsilon_r = 50+5j}$}
	\end{subfigure}	
	\caption{Results for profile shown in Fig. \ref{NumSetup}, we vary the relative permittivity of the scatterer and plot total field magnitude as function of scattering angle. The plots (a)-(h) are respectively for real part of permittivity $\epsilon_R = 1.1, 1.5, 2, 3, 4, 5, 10, 50$ with loss tangent $\delta = 0.1$ which gives relative permittivity $\epsilon_r = \epsilon_R(1+0.1j)$}
	\label{visual_results} 
	\vspace{-0.5\baselineskip}
\end{figure*}


\subsection{Numerical Results and Analysis}
\label{NumExamp}
We first focus on the first profile (shown in Fig. \ref{NumSetup}) and analyze the effect of changing permittivity and loss tangent of the scatterer on the performance of xRA-LM. Later we also investigate the effect of frequency on the performance.  

Fig. \ref{visual_results} provides results for the total received field using xRA-LM, RA, and BA as well as the exact MoM result. The plots of total field in Fig. \ref{visual_results}(a)-(h) are respectively for real part of permittivity $\epsilon_R = 1.1, 1.5, 2, 3, 4, 5, 10, 50$ with loss tangent $\delta = 0.1$ or equivalently, relative permittivity $\epsilon_r = \epsilon_R(1+0.1j)$. 
The relative error (RE) between the estimated and exact field is shown in the legends of each plot.

Fig. \ref{visual_results}(a) shows that for extremely weak scattering ($\epsilon_R=1.1$), both RA and xRA-LM provide comparable performance and are close to the exact field (relative error RE $< 5\%$). BA has large error (RE $=31\%$) which is expected because BA has validity for scatterers smaller in size than $\lambda_0$ whereas the scatterer in this numerical test has diameter $=10 \lambda_0$.

Fig. \ref{visual_results}(b) shows that as the permittivity is increased slightly to $\epsilon_R=1.5$, there is sudden a increase in error (RE $=11\%$) by RA. Whereas xRA-LM still provides accurate estimation with low error (RE $=3\%$). BA again is worse as expected. Fig. \ref{visual_results}(c) shows as permittivity increases to $\epsilon_R=2$, the estimation error of RA increases rapidly (RE$=25\%$) whereas xRA-LM still gives low error (RE $=5\%$). 

Fig. \ref{visual_results}(d)-(f) shows results for permittivity values of $\epsilon_R = 3, 4$ and $5$ respectively. It can be seen that even for these large values of permittivity, our proposed xRA-LM method is able to predict the total field with low error (RE $\le 10 \%$) and outperforms RA and BA by a significantly large margin. Note that these values of relative permittivity are considered very large in the direct/inverse scattering community \cite{chen2018computational, chen2020review, Mittra1998} and to the best of our knowledge, no other non-iterative approximate linear model has been shown to work for this range of permittivity.

In Fig. \ref{visual_results}(g) and Fig. \ref{visual_results}(f), we further increase permittivity to extremely large values of $\epsilon_R = 10$ and $50$ respectively. Even for these extremely strong scattering conditions, xRA-LM provides significantly better performance than RA and BA and acceptable performance in terms of predicting the exact field. The relative error is less than $20\%$ even for extremely large permittivity at $\epsilon_R=10$ which no other non-iterative linear approximation (or even more intricate non-linear models \cite{chen2020review, chen2018computational}) has shown to provide.


		\begin{figure}[!h]
	\captionsetup[subfigure]{aboveskip=2pt,belowskip=5pt, font=footnotesize,oneside,margin={0.4cm,0cm}}
	\centering
	\begin{subfigure}[t]{0.29\textwidth}
		\includegraphics[width=\textwidth]{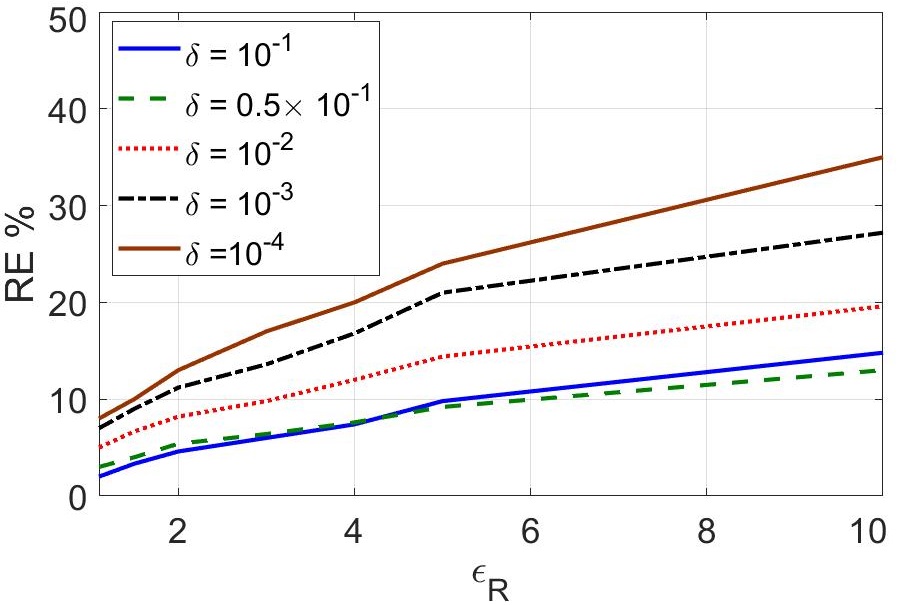}
	\end{subfigure}	
	\caption{Effect of variation of relative permittivity on estimation error for different values of loss tangent $\delta$.}
	\label{summary_plots} 
	\vspace{-0.5\baselineskip}
\end{figure}

Next we investigate the effect of varying loss tangent on the accuracy of xRA-LM. Fig. \ref{summary_plots} provides performance verses loss tangent in the range $\delta \in [10^{-4}, 10^{-1}]$ for fixed value of $\epsilon_R$. We can see that there is increased error if we make loss tangent extremely small. This happens because as the scatterer causes negligible absorption, the higher order reflections inside the boundaries of the scatterer becomes stronger and create stonger multiple scatterer inside the scatterer, and these higher order scattered rays are neglected in our derivation. Fortunately, there is sufficient range of loss tangent $0.001<\delta<0.1$ for which xRA-LM givens acceptable error (RE $< 15\%$) even for extremely strong scattering. This range of loss tangent covers lossy behavior of most of the objects typically found around us (at 2.4 GHz or other microwave frequencies around it)  \cite{4562803, Productnote, ahmad2014partially, Dubey2021}. 

	\begin{figure}[!h]
	\captionsetup[subfigure]{aboveskip=2pt,belowskip=5pt, font=footnotesize,oneside,margin={0.4cm,0cm}}
	\centering
	\begin{subfigure}{0.38\textwidth}
		\includegraphics[width=\textwidth]{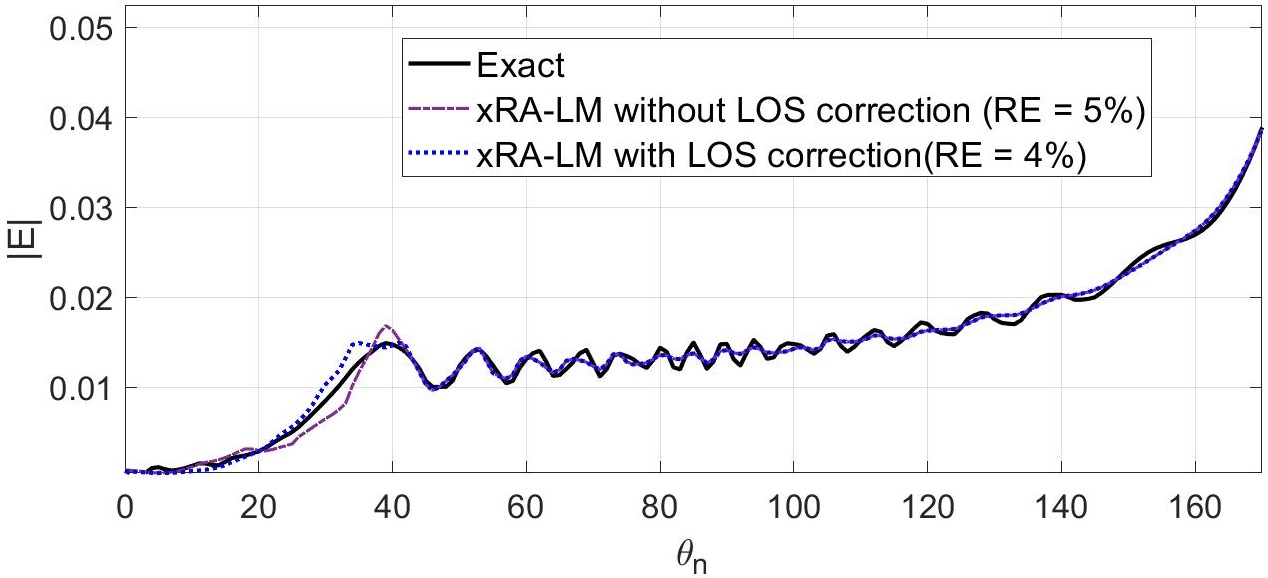}
		\subcaption{$\bm{\epsilon_r=2+j0.2}$}
	\end{subfigure}
	\begin{subfigure}{0.38\textwidth}
		\includegraphics[width=\textwidth]{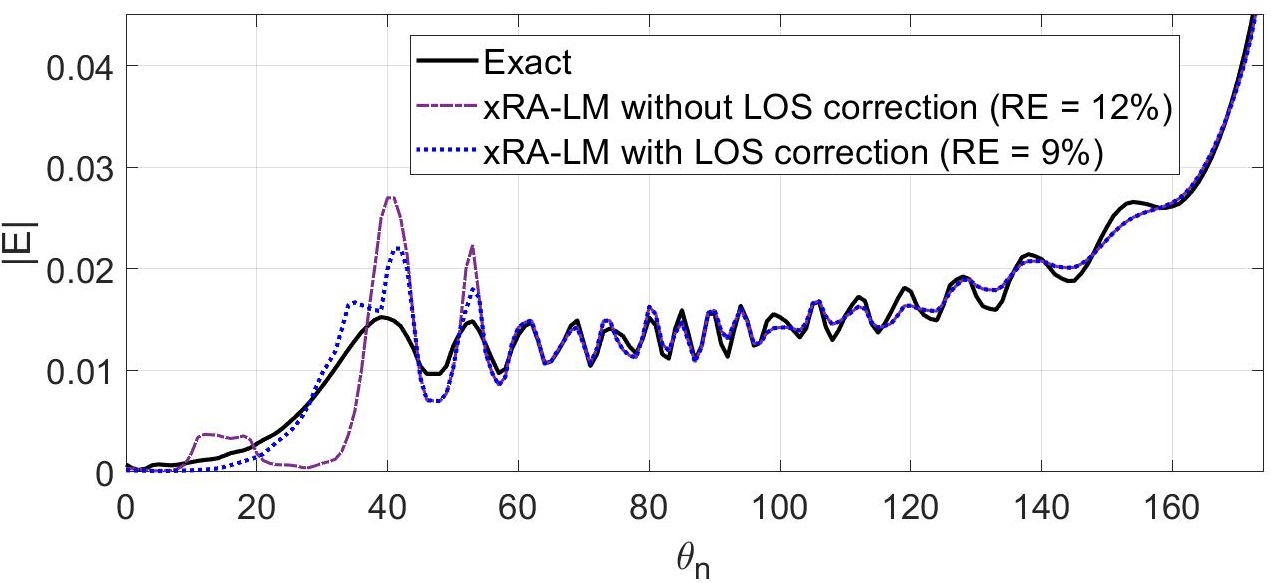}
		\subcaption{$\bm{\epsilon_r=5+j0.5}$}
	\end{subfigure}
	\begin{subfigure}{0.38\textwidth}
		\includegraphics[width=\textwidth]{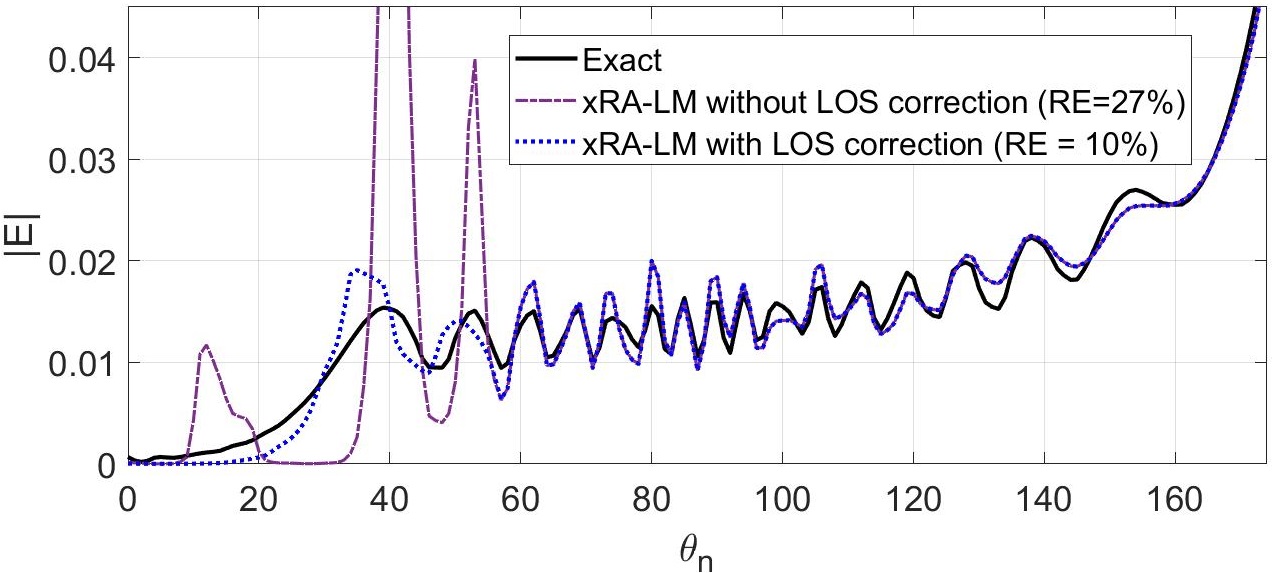}
		\subcaption{$\bm{\epsilon_r=10+j1}$}
	\end{subfigure}	
	\caption{Total field estimation using xRA-LM with and without LOS correction. (a)-(c) shows results respectively for $\epsilon_r = 2+0.2 j, 5+0.5j $ and $10+1j$.}
	\label{visual_results_SK} 
	\vspace{-0.5\baselineskip}
\end{figure}

To provide results for xRA-LM using (\ref{Eq_RIfinal5}) as compared to (\ref{Eq_RIfinal3}), Fig. \ref{visual_results_SK}(a)-(c) provides total field estimation using 1) xRA-LM without the LOS correction (\ref{Eq_RIfinal3}), and 2) xRA-LM with LOS correction (\ref{Eq_RIfinal5}). Fig. \ref{visual_results_SK}(a) shows that for moderate permittivity value of $\epsilon_r=2+0.2j$, xRA-LM provides acceptable error even without LOS corrections. If we further increase permittivity to $\epsilon_r=5+0.5j$ in Fig. \ref{visual_results_SK}(b), accuracy remains good without LOS correction. Finally, in Fig. \ref{visual_results_SK}(c) when permittivity is $\epsilon_r=10+1j$ to create extremely strong scattering, the LOS correction provides significant improvement in accuracy in the forward scattering angles. Whereas for other scattering angles there is no effect of the LOS correction and results are accurate even without it. 

To summarize results from Fig. \ref{visual_results}, Fig. \ref{summary_plots}, and Fig. \ref{visual_results_SK}, the validity range of xRA-LM   (\ref{Eq_RIfinal5}) is in the range $1<\epsilon_R<50$ for all scattering angles.


\begin{figure*}[!h]
	\captionsetup[subfigure]{aboveskip=2pt,belowskip=5pt, font=footnotesize,oneside,margin={0.4cm,0cm}}
	\centering
	\begin{subfigure}[t]{0.38\textwidth}
		\includegraphics[width=\textwidth]{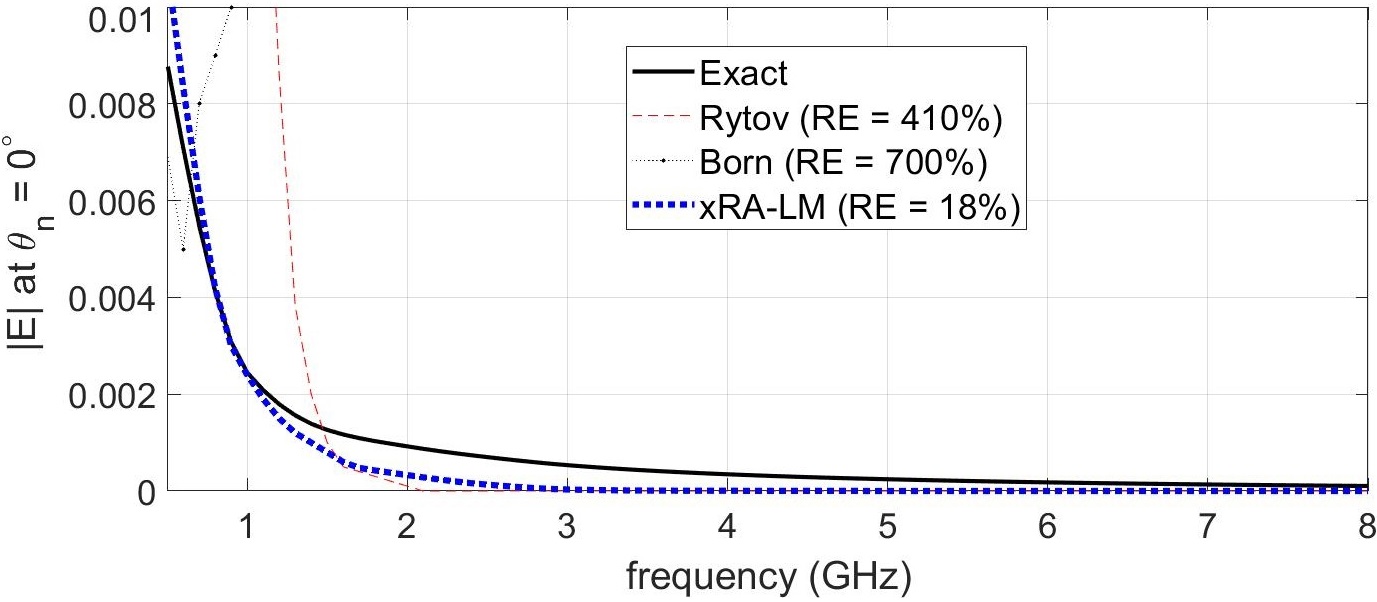}
		\subcaption{$\bm{\theta_n=0^{\circ}}$}
	\end{subfigure}  
	\begin{subfigure}[t]{0.38\textwidth}
		\includegraphics[width=\textwidth]{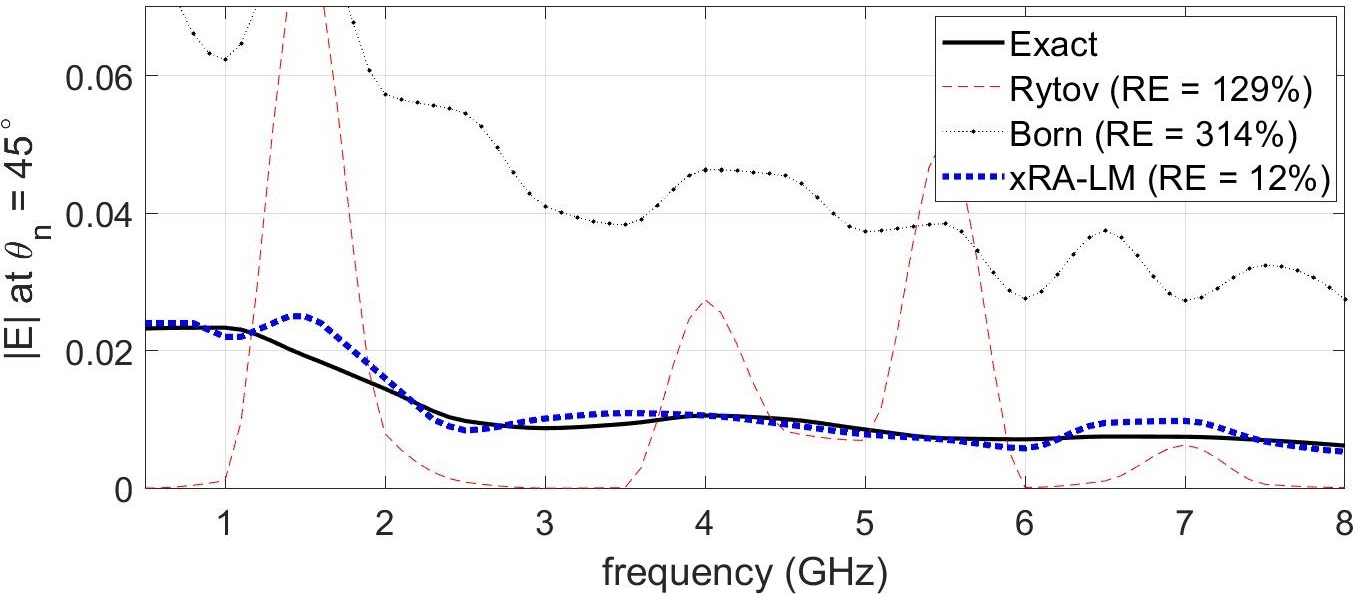}
		\subcaption{$\bm{\theta_n=45^{\circ}}$}
	\end{subfigure}       
	\begin{subfigure}[t]{0.38\textwidth}
		\includegraphics[width=\textwidth]{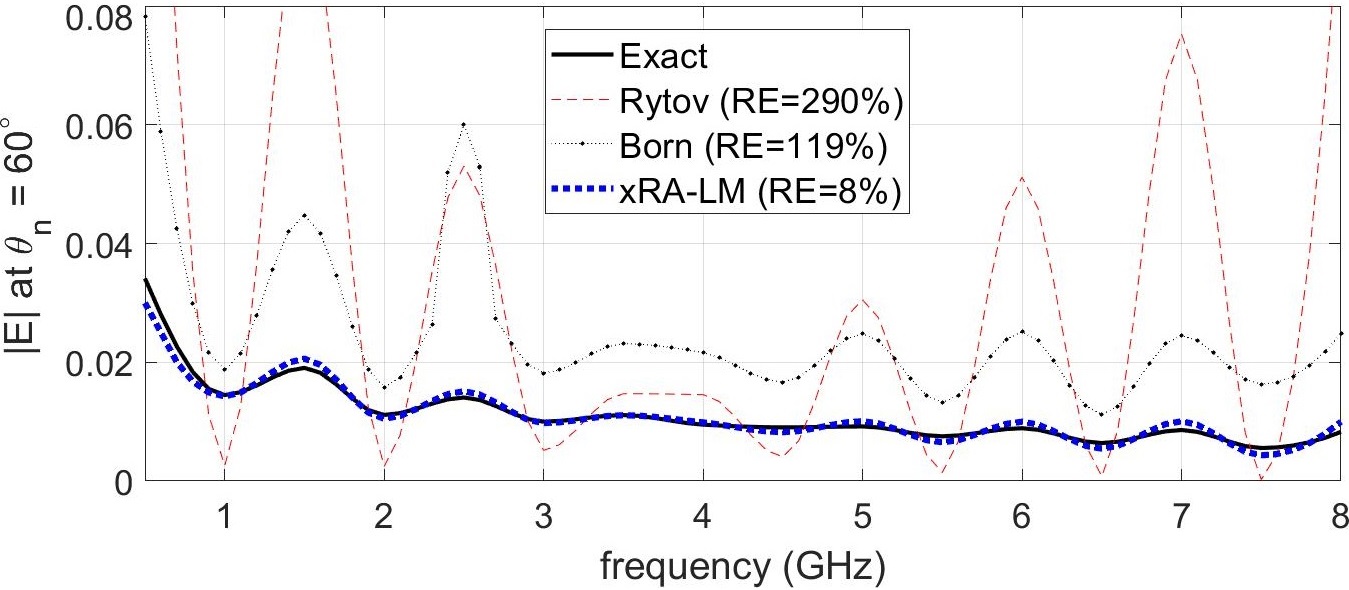}
		\subcaption{$\bm{\theta_n=60^{\circ}}$}
	\end{subfigure}	
	\begin{subfigure}[t]{0.38\textwidth}
		\includegraphics[width=\textwidth]{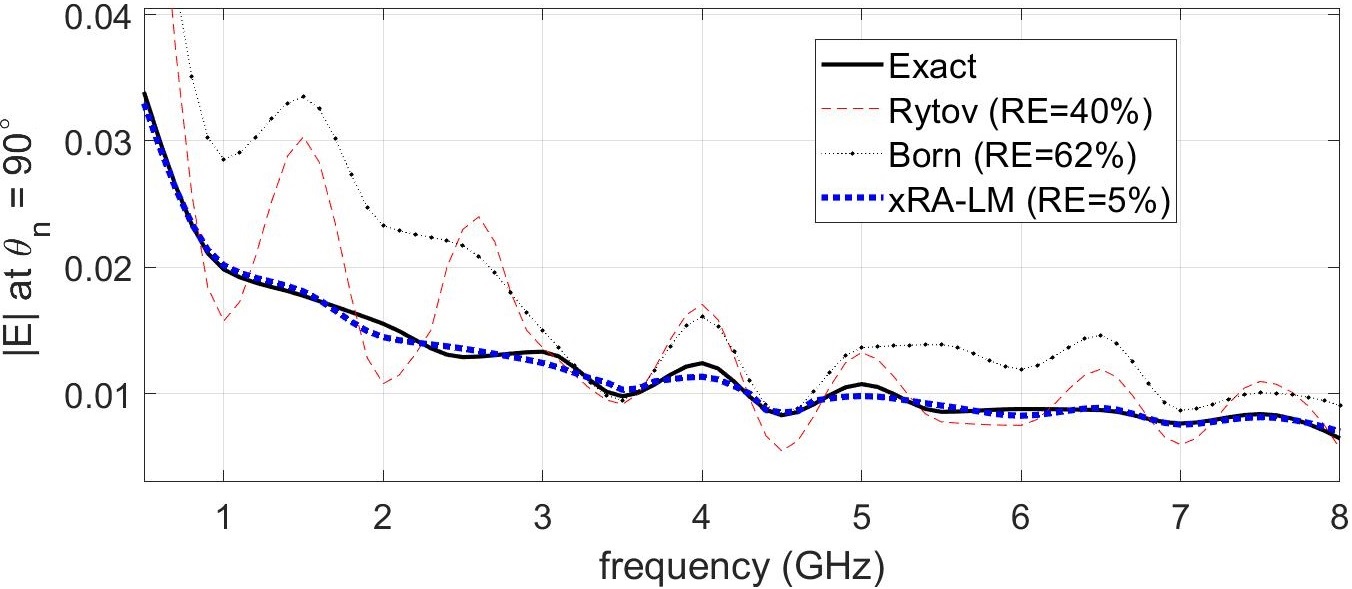}
		\subcaption{$\bm{\theta_n=90^{\circ}}$}
	\end{subfigure} 
	\begin{subfigure}[t]{0.38\textwidth}
		\includegraphics[width=\textwidth]{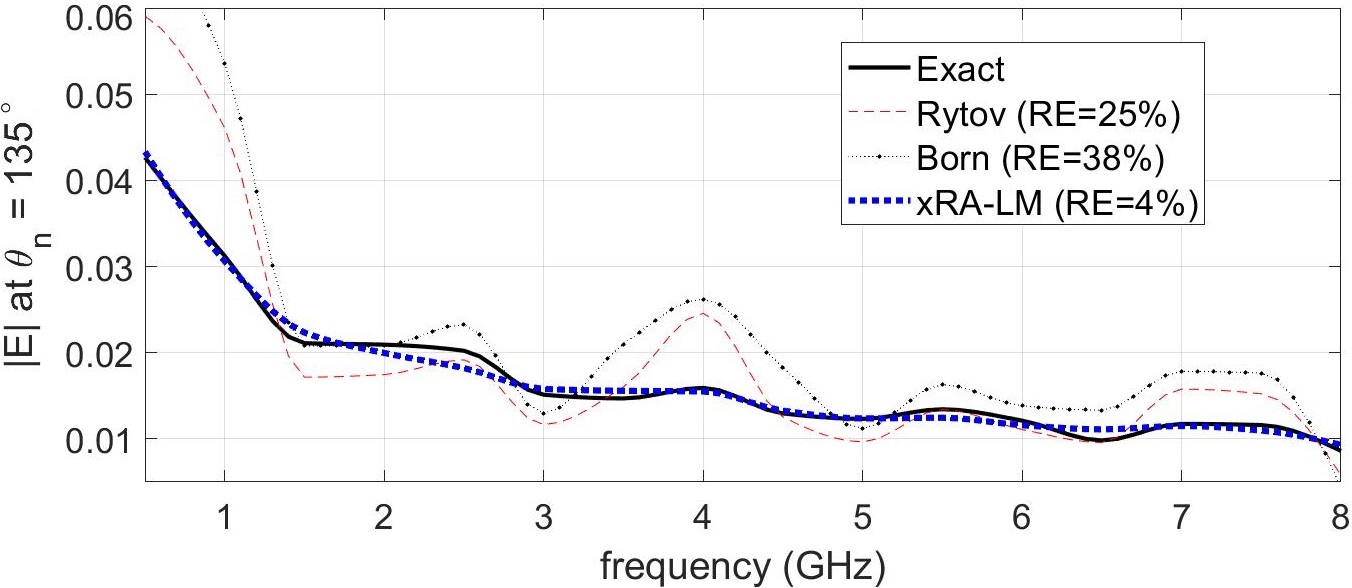}
		\subcaption{$\bm{\theta_n=135^{\circ}}$}
	\end{subfigure}	
	\begin{subfigure}[t]{0.38\textwidth}
		\includegraphics[width=\textwidth]{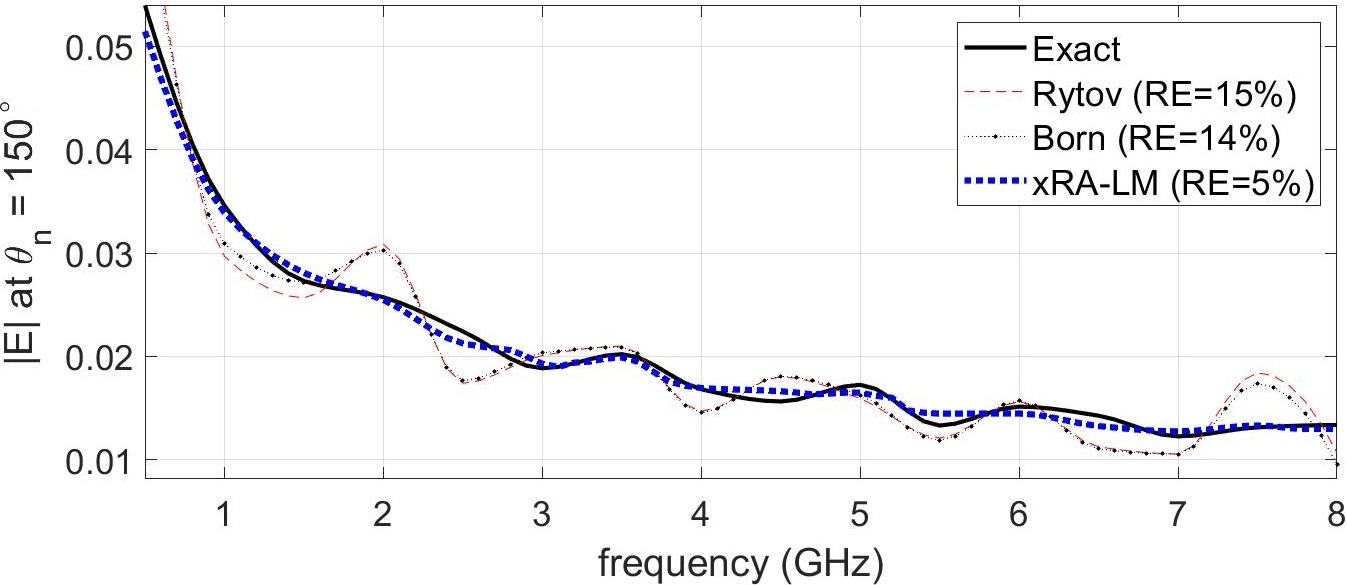}
		\subcaption{$\bm{\theta_n=150^{\circ}}$}
	\end{subfigure}	
	\caption{Effect of variation of frequency on the relative error for estimating total field using xRA-LM, RA and BA. The simulation setup is same as Fig. \ref{NumSetup}. The scatterer size used for estimation is $\epsilon_r = 5 + j 0.5$ to simulate strong scattering conditions. The scatterer is size is 1.25 m. The RE vs. frequency plots in (a), (b), (c), (d) and (e) are for scattering angles $\theta_n = 0^{\circ}, 45^{\circ}, 60^{\circ}, 90^{\circ}$ and $150^{\circ}$}
	\label{multifreq_results} 
	\vspace{-0.5\baselineskip}
\end{figure*}
\begin{figure*}[!h]
	\captionsetup[subfigure]{aboveskip=2pt,belowskip=5pt, font=footnotesize,oneside,margin={0.4cm,0cm}}
	\centering
	\begin{subfigure}[t]{0.37\textwidth}
		\includegraphics[width=\textwidth]{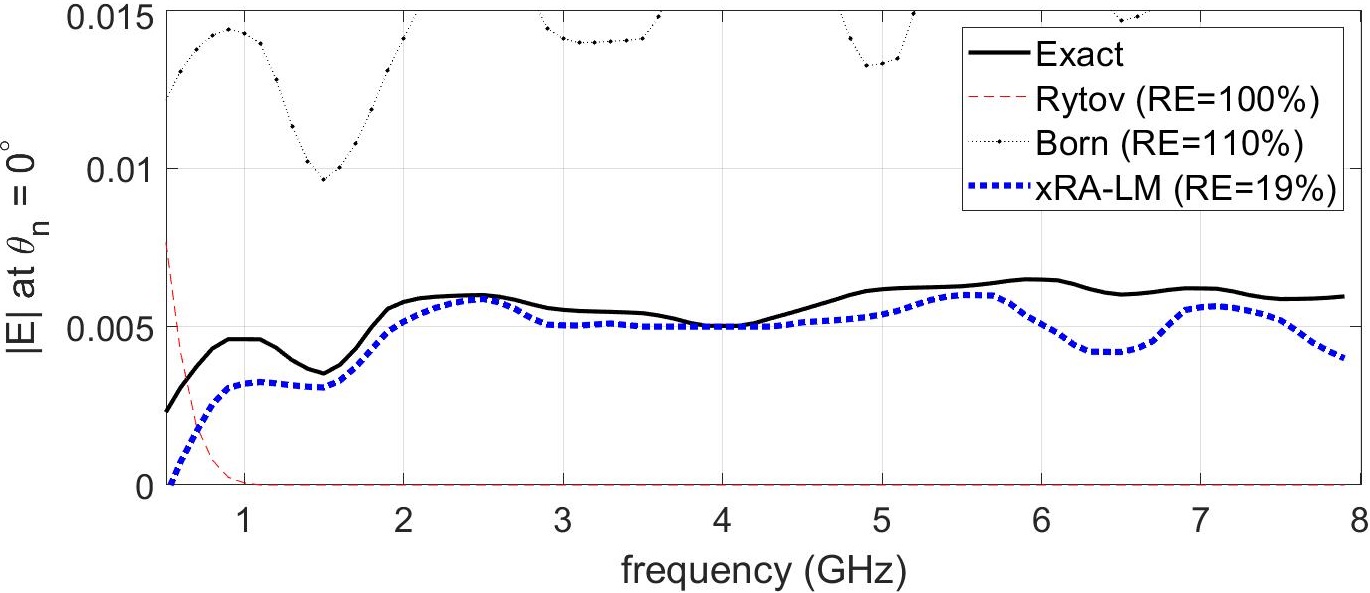}
		\subcaption{$\bm{\theta_n=0^{\circ}}$}
	\end{subfigure}  
	\begin{subfigure}[t]{0.37\textwidth}
		\includegraphics[width=\textwidth]{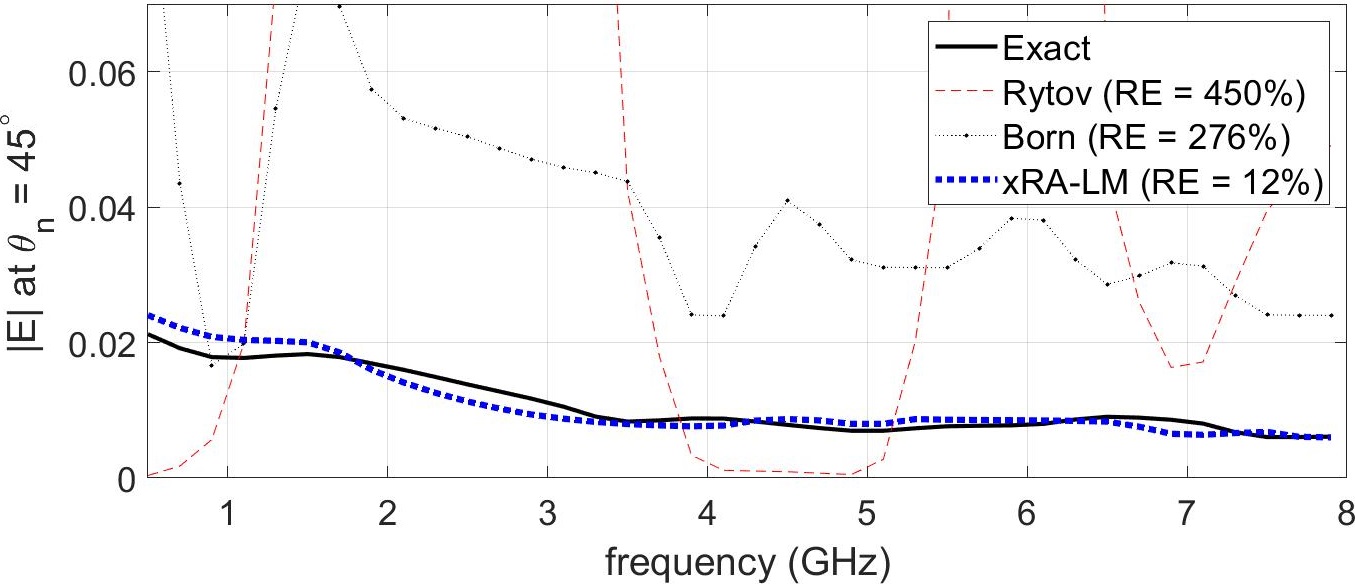}
		\subcaption{$\bm{\theta_n=45^{\circ}}$}
	\end{subfigure}       
	\begin{subfigure}[t]{0.37\textwidth}
		\includegraphics[width=\textwidth]{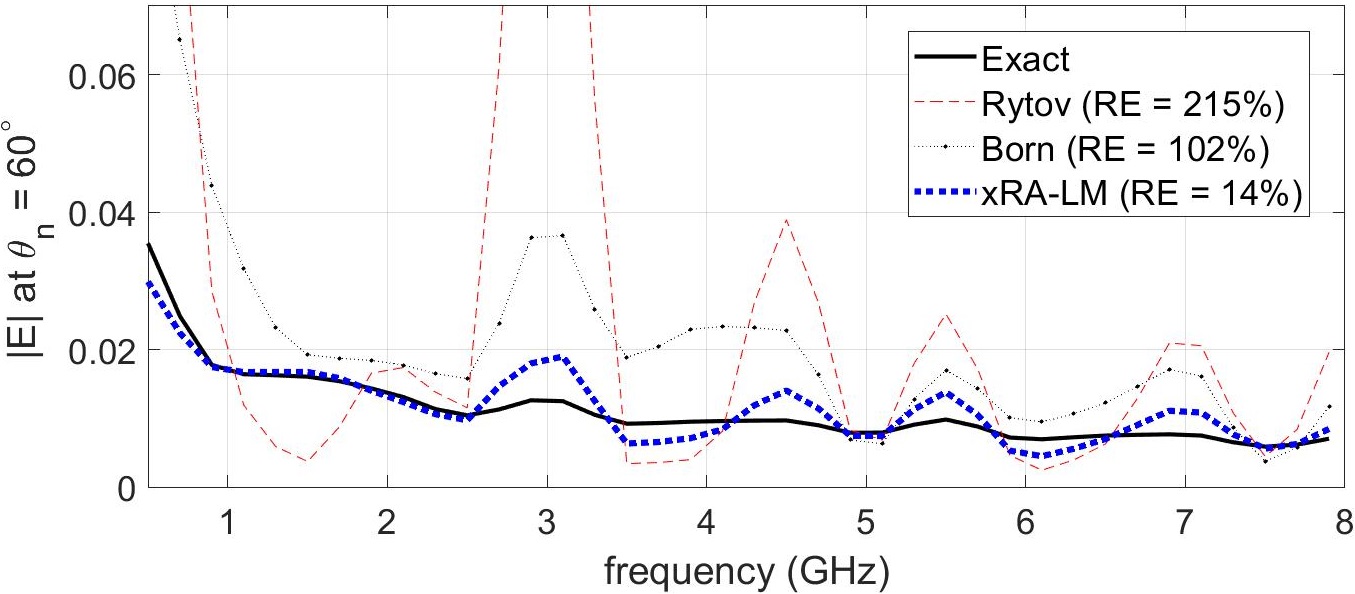}
		\subcaption{$\bm{\theta_n=60^{\circ}}$}
	\end{subfigure}	
	\begin{subfigure}[t]{0.37\textwidth}
		\includegraphics[width=\textwidth]{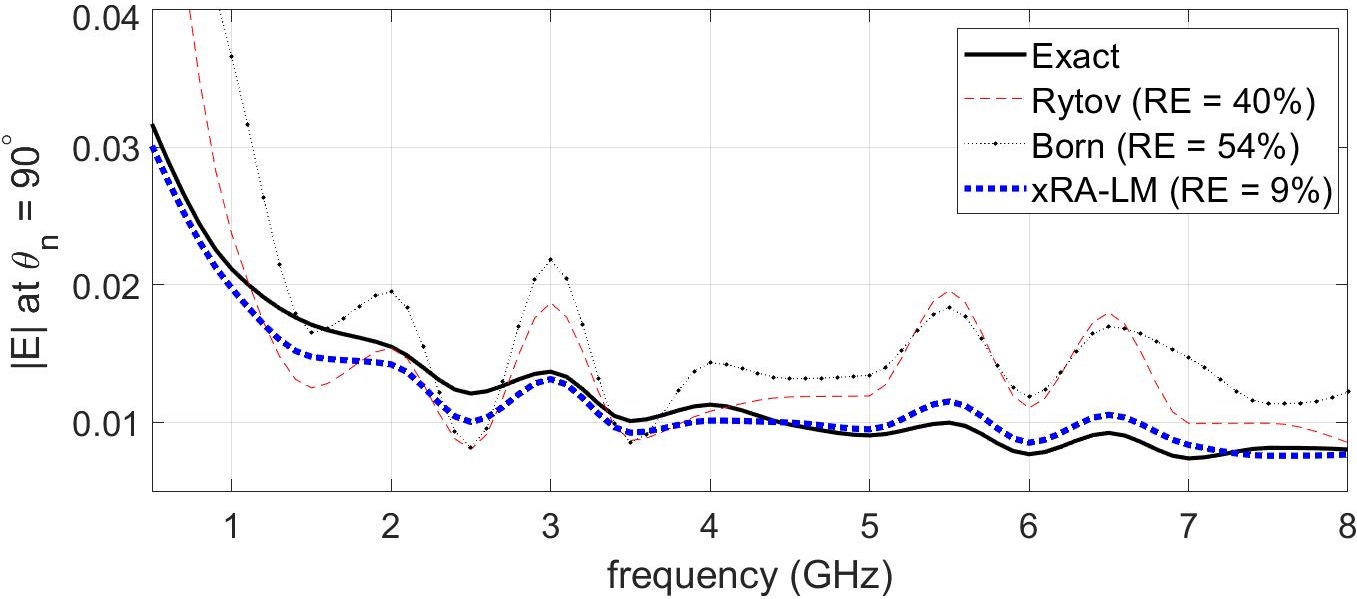}
		\subcaption{$\bm{\theta_n=90^{\circ}}$}
	\end{subfigure} 
	\begin{subfigure}[t]{0.37\textwidth}
		\includegraphics[width=\textwidth]{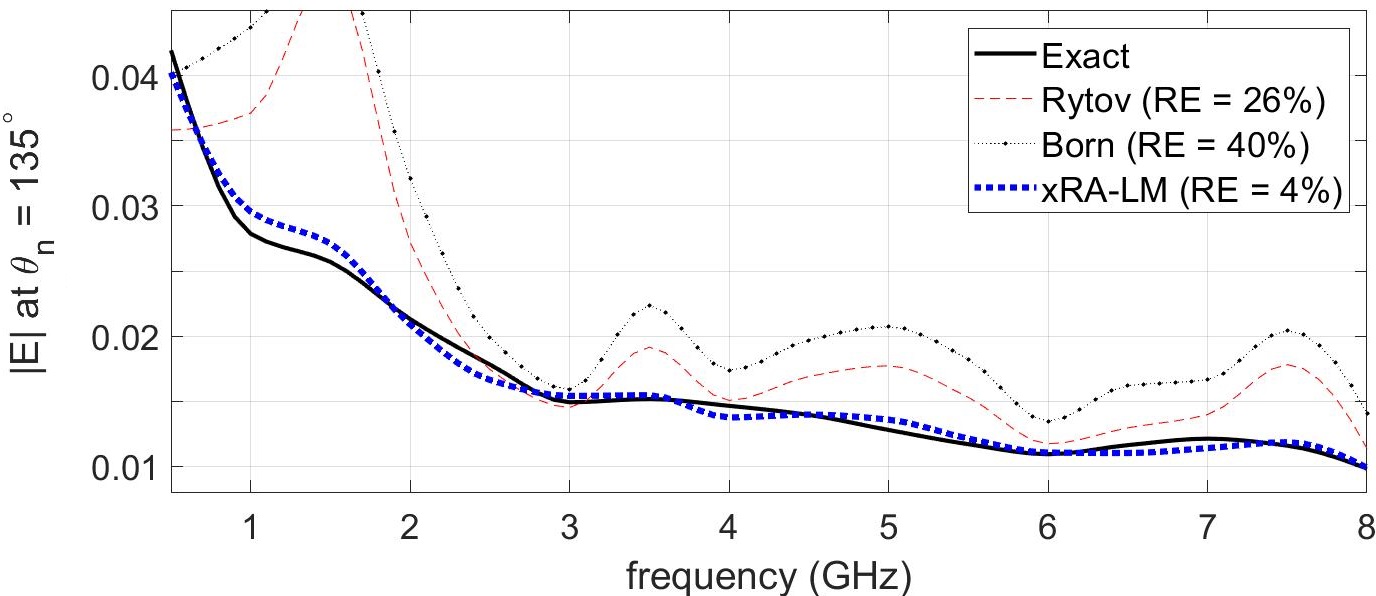}
		\subcaption{$\bm{\theta_n=135^{\circ}}$}
	\end{subfigure}	
	\begin{subfigure}[t]{0.37\textwidth}
		\includegraphics[width=\textwidth]{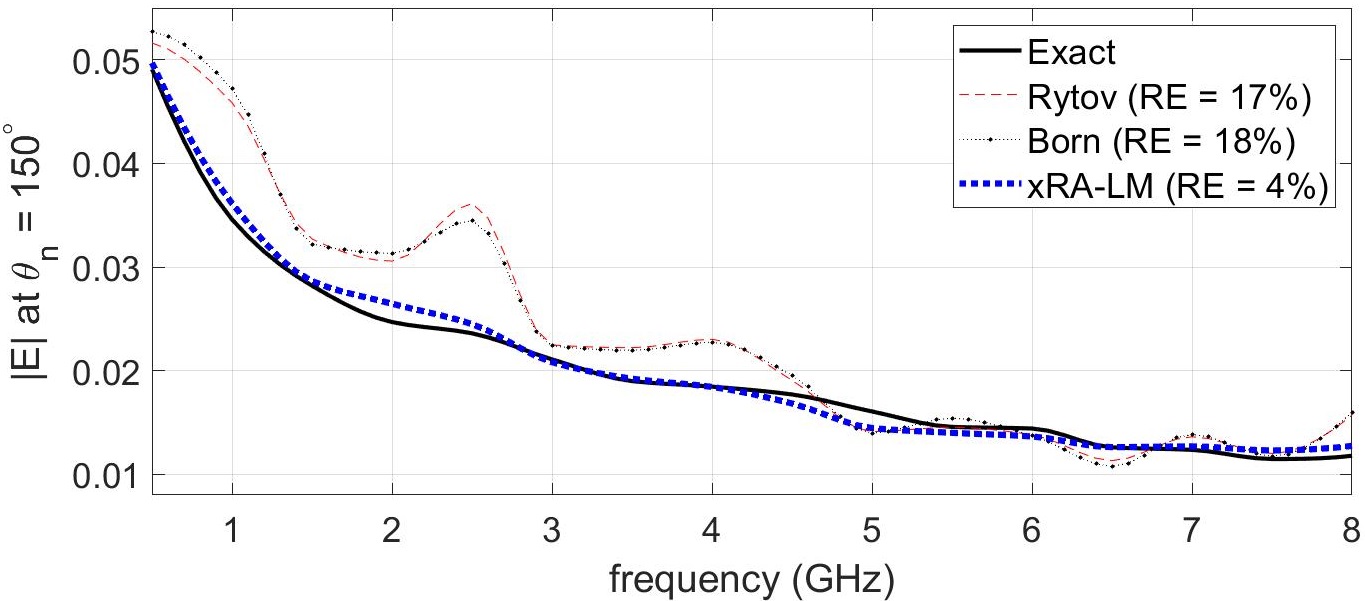}
		\subcaption{$\bm{\theta_n=150^{\circ}}$}
	\end{subfigure}
	\caption{Effect of variation of frequency on the error for multiple scatterer profile in Fig. \ref{NumSetup2}, where circular and square cylinders have permittivity $\epsilon_r = 4+0.4j$ and  $\epsilon_r = 10+1j$ respectively. The RE vs. frequency plots in (a), (b), (c), (d) and (e) are for scattering angles $\theta_n = 0^{\circ}, 45^{\circ}, 60^{\circ}, 90^{\circ}$ and $150^{\circ}$}
	\label{multifreq_results_othershape} 
	\vspace{-0.5\baselineskip}
\end{figure*}

\begin{figure*}[!h]
	\captionsetup[subfigure]{aboveskip=2pt,belowskip=5pt, font=footnotesize,oneside,margin={0.4cm,0cm}}
	\centering
	\begin{subfigure}[t]{0.4\textwidth}
		\includegraphics[width=\textwidth]{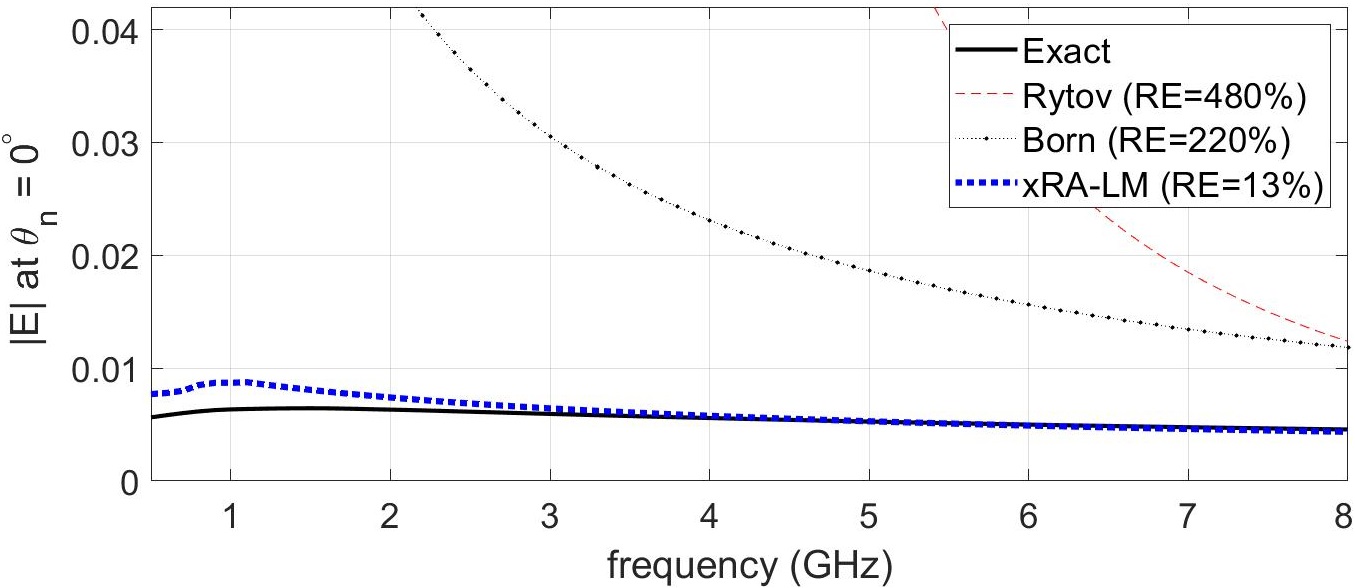}
		\subcaption{$\bm{\theta_n=0^{\circ}}$}
	\end{subfigure}  
	\begin{subfigure}[t]{0.4\textwidth}
		\includegraphics[width=\textwidth]{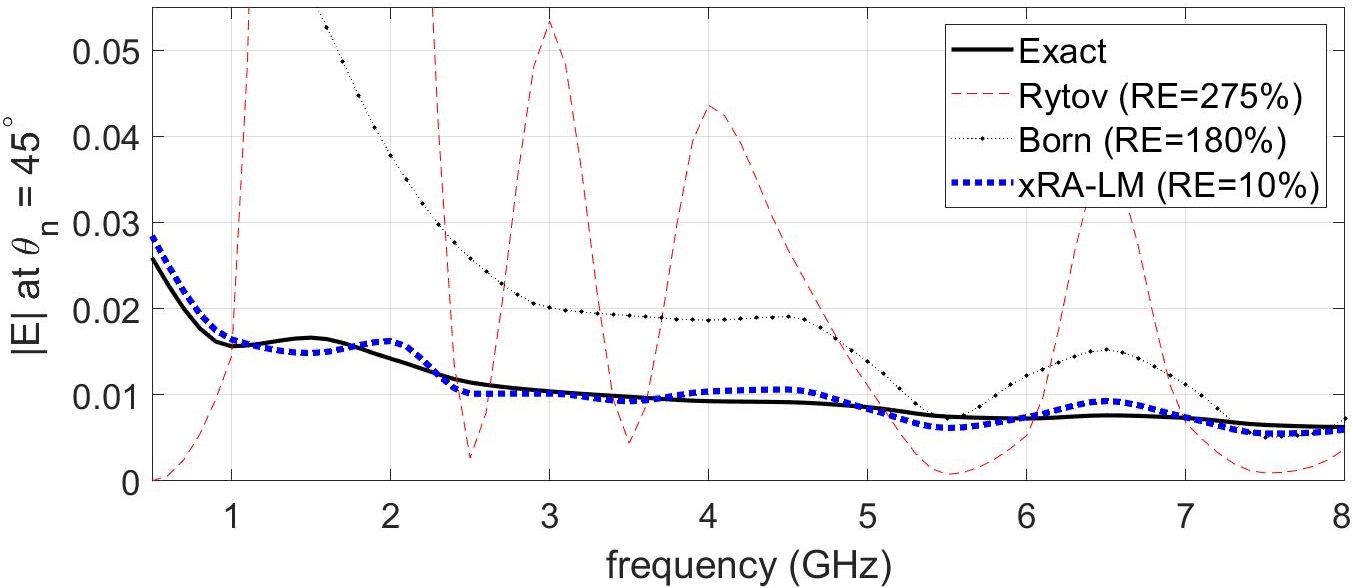}
		\subcaption{$\bm{\theta_n=45^{\circ}}$}
	\end{subfigure}       
	\begin{subfigure}[t]{0.4\textwidth}
		\includegraphics[width=\textwidth]{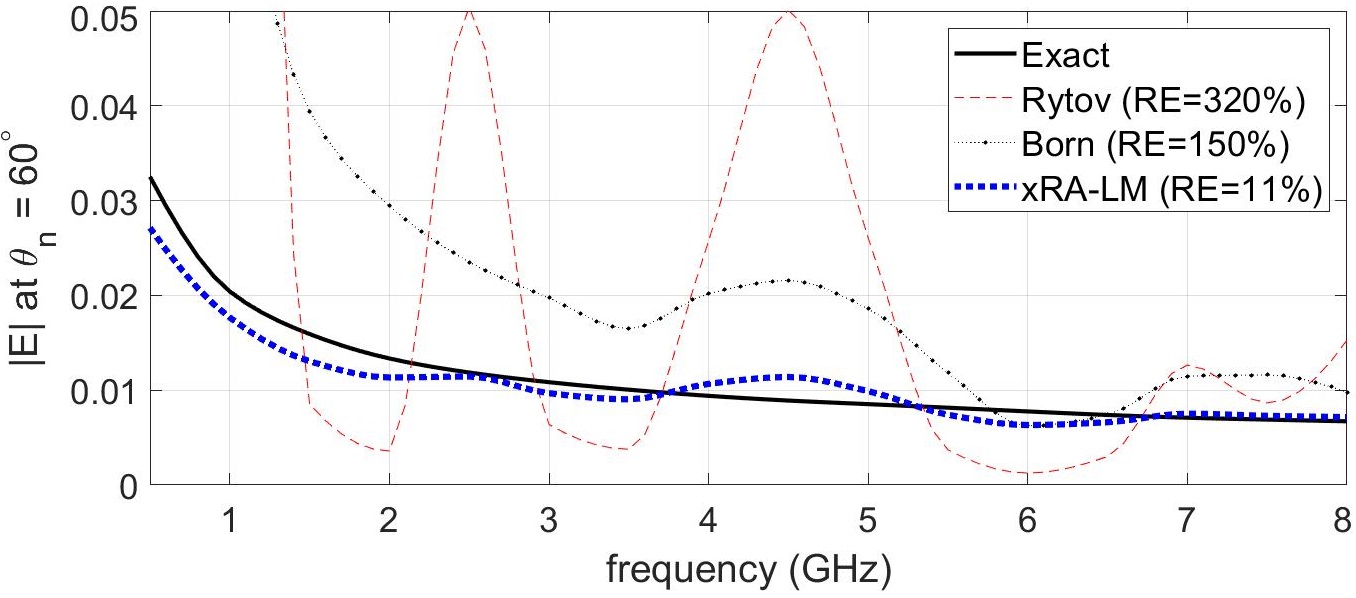}
		\subcaption{$\bm{\theta_n=60^{\circ}}$}
	\end{subfigure}	
	\begin{subfigure}[t]{0.4\textwidth}
		\includegraphics[width=\textwidth]{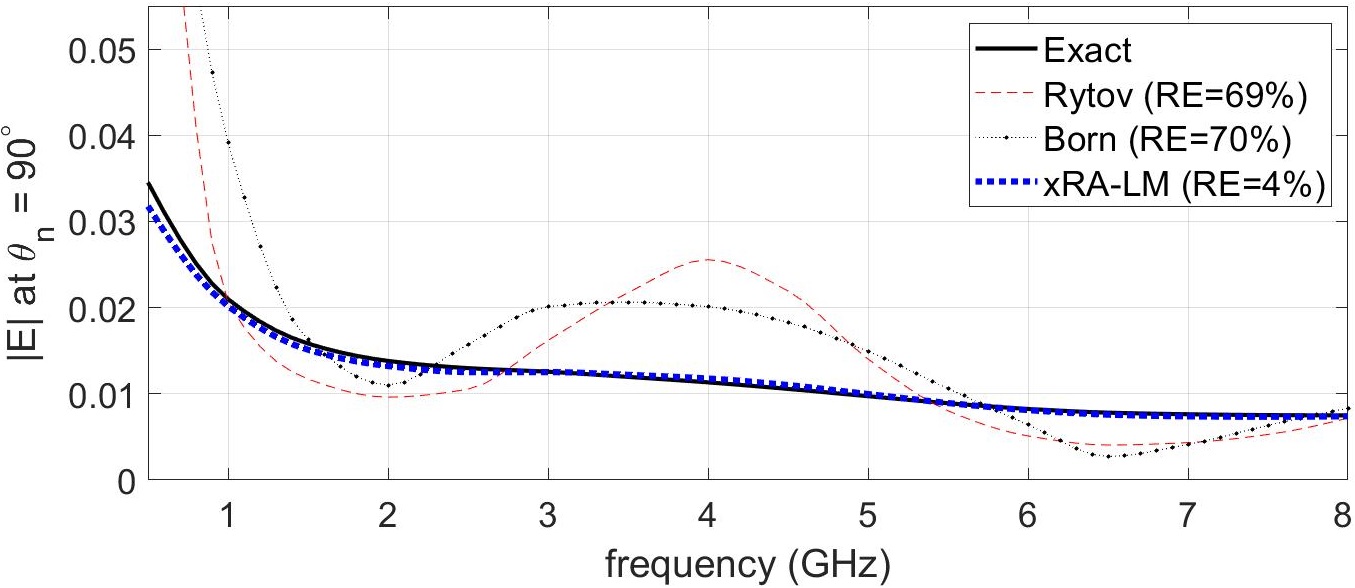}
		\subcaption{$\bm{\theta_n=90^{\circ}}$}
	\end{subfigure} 
	\begin{subfigure}[t]{0.4\textwidth}
		\includegraphics[width=\textwidth]{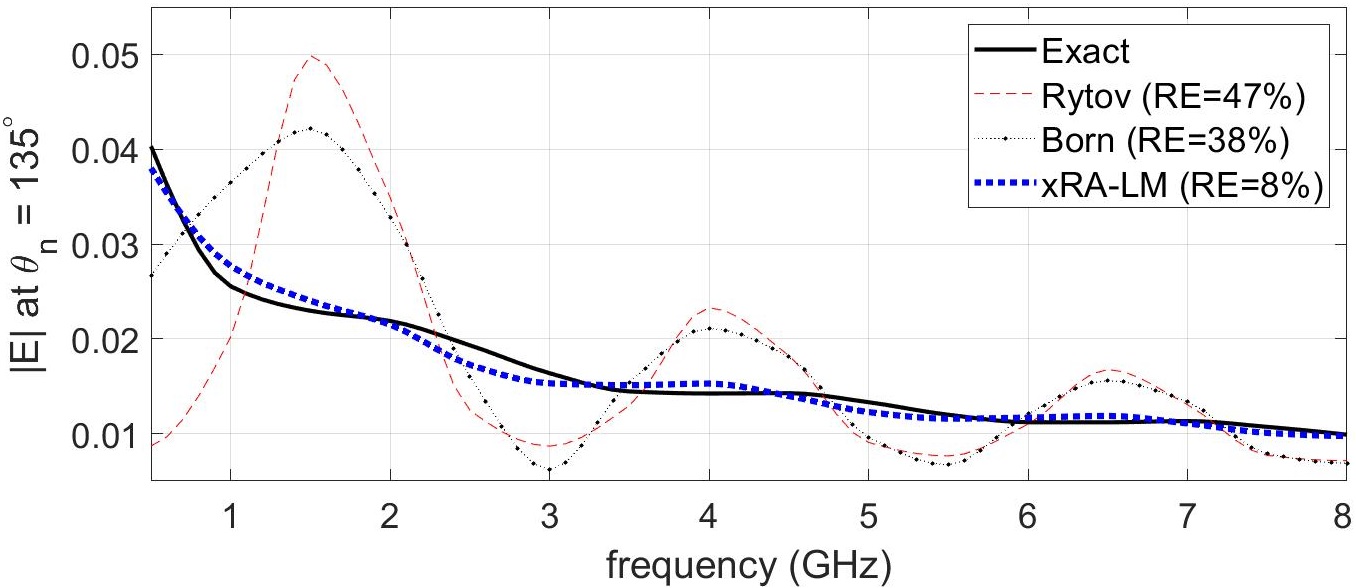}
		\subcaption{$\bm{\theta_n=135^{\circ}}$}
	\end{subfigure}	
	\begin{subfigure}[t]{0.4\textwidth}
		\includegraphics[width=\textwidth]{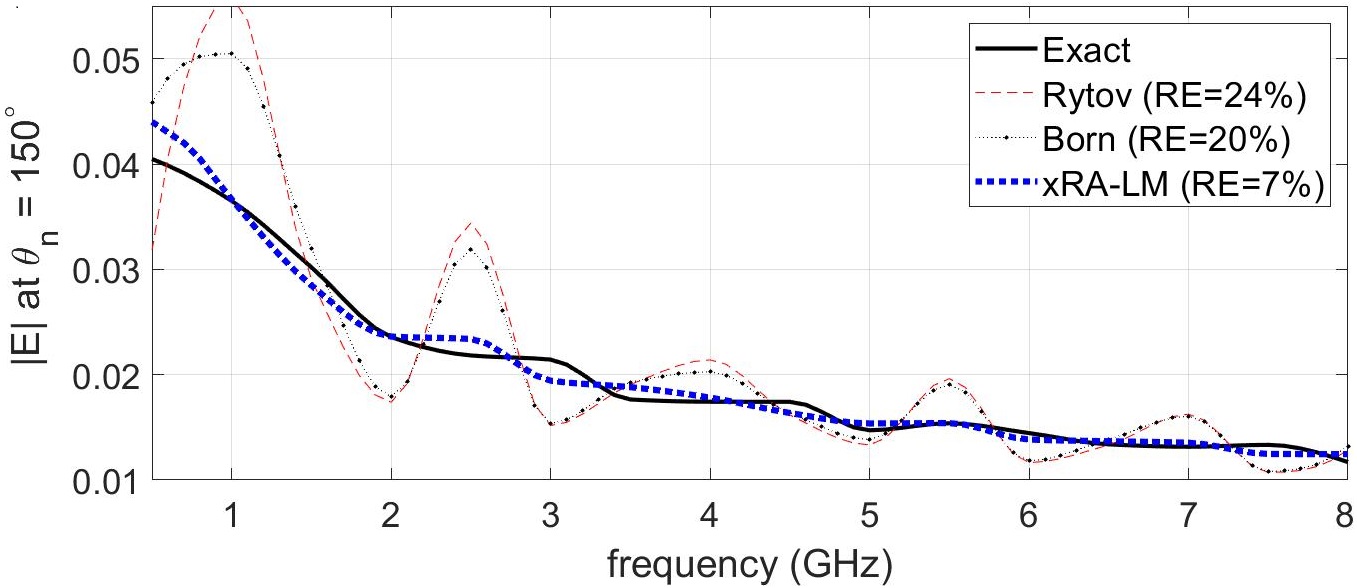}
		\subcaption{$\bm{\theta_n=150^{\circ}}$}
	\end{subfigure}	
	\caption{Effect of variation of frequency on the relative error for Austria profile shown in Fig. \ref{NumSetup3}, where $\epsilon_r = 10+1j$. The RE vs. frequency plots in (a), (b), (c), (d) and (e) are for scattering angles $\theta_n = 0^{\circ}, 45^{\circ}, 60^{\circ}, 90^{\circ}$ and $150^{\circ}$}
	\label{multifreq_results_Austria} 
	\vspace{-0.5\baselineskip}
\end{figure*}

\subsection{Effect of Frequency on Accuracy}
\label{Sec_multifreq}
In this subsection, we investigate the effect of frequency on the accuracy of xRA-LM. (Fig. \ref{multifreq_results}, Fig. \ref{multifreq_results_othershape} and Fig. \ref{multifreq_results_Austria} shows results for profile shown in Fig. \ref{NumSetup}, Fig. \ref{NumSetup2} and Fig. \ref{NumSetup3} respectively). 

The results shown in Fig. \ref{multifreq_results} are for same setup as shown in Fig. \ref{NumSetup}. The permittivity of the scatterer in this test is set to $\epsilon_r = 5 + 0.5j$. However, now we generate results for a range of frequencies rather than only at 2.4 GHz. We vary frequency from $500$ MHz to $8$ GHz (corresponding wavelength varies from $\lambda_0 = 0.6$ m to $0.0375$ m) and show the estimated total field at 6 scattering angles $\theta_n = 0^{\circ}, 45^{\circ}, 60^{\circ}, 90^{\circ}, 135^{\circ}$ and $150^{\circ}$. These scattering angles are selected such that we can see the accuracy of the various methods for forward and back-scattering regions. The results are shown in Fig. \ref{multifreq_results}.

Note that for the largest wavelength of $\lambda_0^\text{max}=0.6$ m, the scatterer diameter ($=1.25$ m) is only twice the incident wavelength and hence the high frequency assumption is not applicable. Whereas at the shortest wavelength of $\lambda_0^\text{min}=0.0375$ m, the scatterer ($1.25$ m) is around $30$ times larger and hence the high frequency assumption is strongly justified. We expect xRA-LM to perform even when the high frequency assumption is not applicable because as we explained in section \ref{Sec_rayRytov}, the cross terms (which are ignored under high frequency assumption in our derivation of corrected contrast) are small due to division by wavenumber $k_0$ and $k_0^2$ terms. Even for the lowest frequency ($\lambda_\text{max} = 0.6$), $k_0 = 10$ and $k_0^2 = 100$ accuracy remains good and cross-talk terms do not appear to add significantly to error. Also, $\nabla \tilde{A}$ terms in the cross-talk terms is non-zero only at the boundaries which further minimizes the cross-talk terms. 


It can be seen in Fig. \ref{multifreq_results} that for all the frequencies and for all scattering angles, xRA-LM provides best estimation of exact field. Even for low frequencies of 500 MHz and 1 GHz, xRA-LM provides low RE ($<13\%$) and significantly outperforms RA and BA. This shows that even when scatterer size is comparable to incident wavelength (high frequency assumption is not strongly imposed), xRA-LM provides good prediction of exact field. This is an important result since it shows that even though xRA-LM is derived by introducing corrections to RA using high frequency approximation, xRA-LM can provide good results even when the high frequency assumption is not strongly present.

Next we provide these results for the multiple scatterer profile (in Fig. \ref{NumSetup2}) and also for the Austria profile (in Fig. \ref{NumSetup3}). Both of these profiles have scatterers of different shapes and different permittivity values and hence provide a good test for strong multiple scattering conditions. The results for the scatterer profile in Fig. \ref{NumSetup2} are shown in Fig. \ref{multifreq_results_othershape} whereas results for the Austria profile (in Fig. \ref{NumSetup3}), are shown in Fig. \ref{multifreq_results_Austria}. These results show that xRA-LM provides highly accurate results and outperform RA and BA by a significant margin, even for these scatterers, for a wide range of frequencies. More broadly, we have found that for all configurations that were considered for the single cylinder, the same conclusions for accuracy can be drawn for the two cylinder and Austria profile configurations.

\section{Conclusion}
\label{Sec_conclusion}
In this paper we present fundamental corrections to the conventional Rytov approximation (RA) using a high frequency approximations for lossy media. It combines the physical interpretation of inhomogeneous wave propagation as rays along-with the diffractive modeling provided by RA. Simulation results demonstrate that for low loss, piece-wise homogeneous scatterers, the proposed xRA-LM method provides good accuracy with errors of 20\% or less even under extremely strong scattering conditions $\epsilon_R = 50$ and outperforms RA and BA by a significant margin. To the best of our knowledge, xRA-LM is the first non-iterative linear approximation for wave scattering that performs well even for extremely large values of permittivity. The technique can open up new paradigms in non-iterative linear approximations that provide feasible solutions to both direct and inverse scattering problems in strong scattering environments with low computational load. It could be particularly important in inverse scattering contexts where the formulations are inherently non-linear and ill-posed and the measurement data suffers from inaccuracies due to noise and real world data acquisition.


\bibliographystyle{IEEEtran}
\bibliography{TAP-v16}
%
\end{document}